\documentclass[twocolumn, superscriptaddress,groupedaddress]{revtex4}

\usepackage{graphicx}  % needed for figures
\usepackage{dcolumn}   % needed for some tables
\usepackage{bm}        % for math
\usepackage{amssymb}   % for math

\usepackage{mathtools}

\usepackage[T1]{fontenc}
\usepackage[utf8]{inputenc}
\usepackage[english]{babel}

\usepackage[nolist,nohyperlinks]{acronym}

\begin{document}

\title{\textit{Ab initio} RNA folding}
\author{Tristan Cragnolini}
\affiliation{Laboratoire de Biochimie Th\'eorique UPR 9080 CNRS, Universit\'e Paris Diderot, Sorbonne, Paris Cit\'e, IBPC
13 rue Pierre et Marie Curie, 75005 Paris, France}
\affiliation{now at University Chemical Laboratories, University of Cambridge, Lensfield Road, Cambridge CB2 1EW, United Kingdom}

\author{Philippe Derreumaux}
\affiliation{Laboratoire de Biochimie Th\'eorique UPR 9080 CNRS, Universit\'e Paris Diderot, Sorbonne, Paris Cit\'e, IBPC
13 rue Pierre et Marie Curie, 75005 Paris, France}
\affiliation{Institut Universitaire de France, Boulevard Saint-Michel, 75005, Paris}
\author{Samuela Pasquali  \footnote{Corresponding author : samuela.pasquali@ibpc.fr}}
\affiliation{Laboratoire de Biochimie Th\'eorique UPR 9080 CNRS, Universit\'e Paris Diderot, Sorbonne, Paris Cit\'e, IBPC
13 rue Pierre et Marie Curie, 75005 Paris, France}

\begin{abstract}
RNA molecules are essential cellular machines performing a wide variety of functions for which a specific three-dimensional structure is required.
Over the last several years, experimental determination of RNA structures through X-ray crystallography and NMR seems to have reached a plateau in the number of structures resolved each year, but as more and more RNA sequences are being discovered, need for structure prediction tools to complement experimental data is strong.
Theoretical approaches to RNA folding have been developed since the late nineties when the first algorithms for secondary structure prediction appeared.
Over the last 10 years a number of prediction methods for 3D structures have been developed, first based on bioinformatics and data-mining, and more recently based on a coarse-grained physical representation of the systems.
In this review we are going to present the challenges of RNA structure prediction and the main ideas behind bioinformatic approaches and physics-based approaches.
We will focus on the description of the more recent physics-based phenomenological models and on how they are built to include the specificity of the interactions of RNA bases, whose role is critical in folding.
Through examples from different models, we will point out the strengths of physics-based approaches, which are able not only to predict equilibrium structures, but also to investigate dynamical and thermodynamical behavior, and the open challenges to include more key interactions ruling RNA folding.
\end{abstract}

\maketitle

\tableofcontents

\section{Introduction}
Over the last fifteen years it has been recognized that RNAs play a wide range of functions aside for their well known roles of genetic information carrier (mRNA) and amino acid recruiter (tRNA):
microRNA (miRNA) are short sequences regulating genes in the post-transcriptional process, the small interference RNA (RNAi) acts on the gene silencing mechanism,
ribozymes are mid sized (often less than 100 nucleotides) molecules with catalytic properties,
and ribosomal RNA constitutes the ribosome together with proteins and can be as large as several thousands nucleotides.
The publication of high-resolution X-ray structures revealed that the catalytic activity
in the ribosome was carried by RNA, and not the associated proteins\cite{Ban2000,Nissen2000}.
Many more ribozymes have been identified. The RNase P is necessary for the maturation of tRNA,
while intron splicing is catalysed by a protein-RNA complex, the spliceosome\cite{Toor2008}.
Other ribozymes play role in metabolic pathways, such as the glucosamine-6-phosphate (glmS) ribozyme,
regulating the translation of the protein catalysing the production of glmS\cite{Ferre-D'Amare2010}.
There is also growing interest in the use of RNA for nanotechnology, with the creation of self-assembling systems such as artificial nanorings\cite{Grabow2011},
nanocages\cite{Hao2014}, nanoscale scaffolds\cite{Afonin2010,Khisamutdinov2014} and other nanostructures\cite{Jaeger2006}.
More recently, riboswitches have been identified. Those sequences, usually present in the 5' untranslated region of genes, adopt a specific fold in the presence or absence of a ligand.
The folding of the riboswitch sequence will then regulate the expression of the associated gene.
Riboswitches have been identified for purine bases, adenine and guanine, for amino acids, notably tryptophan,
as well as organic compounds, such as fluoride.
RNA is also a prime candidate for being a key molecule in the emergence of life on earth\cite{Obermayer2011},
Like proteins, the functionality of these molecules depends crucially on their equilibrium structures and their dynamical behavior \cite{Holbrook2005, Strobel2008}, 
with distinct active conformations biologically active under different conditions \cite{Scott2007}.
This poses the problem of understanding RNA folding, that is why and how a specific sequence adopts a specific tertiary structure.

The ENCODE project showed that a large number of non-protein-coding RNA transcripts were produced,
most of them with no previously recognised roles\cite{Birney2007a}.
With the explosion of sequencing data,
with nearly 200 millions entries contributed to GenBank over the last 30 years,%(figure~\ref{db_count}),
and most DNA being detected as ``non-coding'',
therefore possibly containing the information to synthesize RNAs, structure prediction from sequence is an urgent matter. 
High resolution experimental techniques for determining three-dimensional structures, such as X-ray crystallography and NMR,
are challenging as it is shown by the small number of resolved structures in the 
Nucleic Acids Data Bank (NDB) and by the scarcity of structures with substantially different architectures. 
Low-resolution techniques, such as SAXS and Cryo-EM, allow for easier access to the raw data, but require extensive modeling to propose a well-resolved structure.

\subsection{RNA structural organization}
Before entering the details of RNA folding predictions it is useful to outline the different levels of complexity that are involved.
RNA, just like DNA, benefits from sequence complementarity, with A pairing with U and G pairing with C.
If we have strands with perfect complementary sequences, the structure of the molecule is a perfect helix (for RNA an A-form).
In this case predicting the fold of the molecule is rather trivial as the characteristics of the helix (rise, pitch, ...) are well known.
But RNA sequences almost never allow for base complementarity along the whole sequence.
RNAs are most often single stranded molecules that have sequences incompatible with the formation of long double helices.
Nonetheless they can have short portions of complementary sequences giving rise to short helices separated by single stranded regions.
Portion of the sequence close by tend to form helices and give rise to hairpins, with a helical stem and a terminating loop of variable size.
Helices and single stranded regions arrange in space with the possible formation of base pairs external to helices. 
Often these contacts exhibit non-canonical pairings, that is base pairs other than AU or CG, and involve all sides of the base \cite{Leontis2002b}.

\begin{figure}[tb]
\begin{center}
\includegraphics[width=0.45\textwidth]{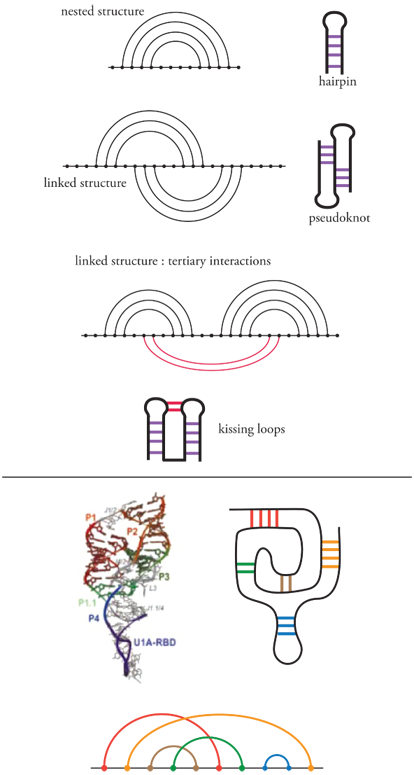}
\caption{{\small Top: schematic representation of base-pairing organization to give nested structures and hairpins, and non-nested structures resulting in pseudoknots and other tertiary interactions. Bottom: 2D and 3D structure of the HDV ribozyme (PDB code 1DRZ). The bottom graph shows the intricate base-pairing connections forming the various helices (color coded).}}\label{fig-1}
\end{center}
\end{figure}

If for proteins the definition of secondary and tertiary structure comes unambiguously from backbone hydrogen bonds,
for RNA the definition is more delicate because base pairing occurs both at intermediate lengths scales with hairpins,
and at large length scales with bonds holding together already formed structures.
We adopt the following definition:
two base pairs I $\circ$ J, H $\circ$ K, are called nested if I$<$H$<$K$<$J, unrelated if I$<$J$<$H$<$K, linked if I$<$H$<$J$<$K.
RNA {\it secondary structure} is a set of base pairs in which no two base pairs are linked, that is, every base pair in a secondary structure is either nested or unrelated \cite{Zuker2000a}. 
If we represent secondary structures as graphs with nucleotides identified with nodes and lines representing base pairs, this is equivalent to saying that secondary structures can be represented 
by planar graphs in which no lines intersect (Figure \ref{fig-1}).
We can classify basic secondary structures into single stranded regions, hairpins, bulge loops, mismatches, internal loops, and junctions \cite{Chastain1991}.

Nucleotides which are linked form {\it tertiary interactions}.
Most tertiary interactions involve non-canonical base pairing or backbone-backbone interaction. 
Some tertiary structures are adenosine platforms, triplets, helices docking, metal-core motifs, ribose zippers. 
Adenosine platforms and triples involve multiple base pairing, helix docking involve non-canonical pairing as well as backbone-backbone or backbone-base interactions, 
ribose zippers involve backbone-backbone interactions.
Tertiary structures involving canonical pairing are pseudoknots and kissing loops. 
Pseudoknots and loop-loop interactions tie together single stranded regions. \\
The definition of tertiary structure adopted for RNA is clearly different from the definition commonly used for proteins, where the term refers to the global organization of 
secondary structure elements in space.
For RNA we will refer to the three-dimensional global organization as to the {\it architecture}.

Early experiments on RNA melting showed that RNA unfolds in a series of discrete steps corresponding to the breaking down of the folding process into localized regions of the structure \cite{Tinoco1999}.
More recent analyses on large ribosomal molecules show that RNA has a rich modular structure \cite{Holbrook2005, Reiter2011}, findings also supported by numerous single molecule 
pulling experiments \cite{Onoa2004, Li2008a}.
These findings suggest that an RNA molecule possesses a hierarchical structure in which the primary sequence determines the secondary structure which in turn determines its tertiary folding: the three-dimensional architecture results from the compaction of separate pre-existing and stable elements that form autonomous entities.
Exceptions to this general scheme exist, as it is the case for some complex architectures and pseudoknots, where melting of tertiary structures are not well separated form melting of secondary 
structures \cite{Gluick1994}.
Because of the stability of base-pairing at room temperature and of the frustration in the possible secondary structures, RNA can adopt dramatically different conformations all of similar energetic stabilities.
Some of these structures have indeed alternative conformations that the molecule can adopt in response to environmental conditions,
others are kinetic traps that can lead to the molecule degradation by regulating factors \cite{Al-Hashimi2008, Serganov2009,Fuertig2007}.
Typically small RNA molecules reach their native state without being trapped in misfolded structures, while long molecules are trapped more easily with increasing chain length \cite{Pan1997}. 

In conclusion, despite what one could naively think based on base complementarity, the folding of RNA can not be considered fully as a hierarchical problem.
Whether secondary and tertiary structures can be treated separately depends on the molecule's size and structural complexity.

\subsection{Computational challenges}
With the recent access to massive computational resources and with the establishment of reliable atomistic force fields, one would think that numerical studies of RNA folding should not pose a problem. 
In the late nineties the common belief was that if only a fraction of the resources put in solving protein folding was put into RNA folding, the problem would have been solved quickly \cite{Tinoco1999}.
Yet, fifteen years later, despite the increase of human and computational resources, the question is still open.
The obvious approach using an atomistic model and simulations proves particularly challenging, and in practice limits the studies to very small molecules and short times \cite{Chen2013, Sorin2004}.
The main difficulty comes from size of the molecule and time of structure formation, as several different lengths scales are involved for both.
When RNA renature, stem-loops form in microseconds, while global architectures can take seconds to minutes do develop. 
Even if we were able to simulate efficiently at the lowest time scale, which is far from being the case with atomistic simulations, there would be several orders of magnitude 
in time to look at for just one folding event, not to mention any statistical analysis.
Concerning size, even just at the level of secondary structure, molecules with more than a dozen nucleotides have a multitude of possible states and base-pairing space is quickly extremely large \cite{Gan2003}, with different structures separated by large energy barriers.
Two additional problems come from the high charge of the RNA backbone, giving rise to crucial interactions with both solvent and ions in solution, and the intrinsic nature of 
hydrogen bonding and stacking that would require quantum mechanical calculations for accurate results \cite{Sponer2008}.
Because of these challenges, nucleic acids force fields are still far from being as reliable as one would like them to be.
AMBER, which among classical atomistic force fields is the one that has been developed more carefully for nucleic acids, works well for small helical structures, but 
for admission of its own developers, fails in the study of RNA single stranded molecules, as the configurational changes involved go beyond the testing ground of its parameters \cite{Cheatham2013}.

Given the limitations of atomistic simulations, different strategies have been applied to RNA structure prediction and can be loosely organized into three categories \cite{Rother2011}: knowledge-based homology models, hybrid bioinformatic methods, and coarse-grained \textit{ab initio} models.

As it is the case for proteins, when the question is that of determining a 3D configuration with the best possible accuracy, homology models based on sequence similarity perform well, provided one can find an already resolved structure that serves as template \cite{Rother2011a,Flores2010}. This is rarely the case for single stranded RNA.\\
The hybrid category comprise a large variety of methods, going from fragment reconstructions \cite{Das2007,Parisien2008} to models strongly relying on secondary structure prediction algorithms and 3D scaffolds extracted from the NDB \cite{Cao2011}. 
In general these methods are good in predicting local structures, but have their weakness in the prediction of overall complex architectures, unless experimental additional constraints on the tertiary structure are known.\\
Coarse-grained \textit{ab initio} models try to capture the physics of the system, and aim at predicting equilibrium structures as well as folding intermediates and energy landscapes. 
Atoms are grouped in particles constituting the elementary objects of the model, and a set of forces are defined to generate a dynamic.

In this review we will present the basic principles of the different strategies of physics-based models for RNA folding.
To put these developments into context, in section \ref{sec_atom}  we will discuss the state of the art of single stranded nucleic acids atomistic simulations and in
in section \ref{sec_bioinfo} we will discuss bioinformatic approaches, with the explicit examples of FARNA/FARFAR \cite{Das2007}, Vfold \cite{Cao2011}, MC-fold \cite{Parisien2008}.
Section \ref{sec_CG} constitutes the main body of the review and focuses on \textit{ab initio} coarse-grained models.
Here we will address the various issues going into building a CG model for RNA : choice of particles to represent the system, 
choice of the functions describing the interactions, parametrization.
These aspects will be developed in detail for the models that currently have the best prediction capabilities, namely the model by Xia \cite{Xia2010}, OxRNA \cite{Sulc2014a} and HiRE-RNA \cite{Cragnolini2014}.
We will also discuss the simulation methods employed by coarse-grained models for structure prediction (section \ref{sec_sim}) and discuss the performance of both bioinformatics and \textit{ab initio} approaches on some benchmark systems (section \ref{sec_bench}).
Section \ref{sec_exp} presents how coarse-grained models can be coupled to existing experimental data to obtain structures fulfilling constraints coming from experiments.
We will conclude with a discussion on the open challenges coming from the interplay of ions and RNA and the interactions of bases with other groups, such as phosphates (section \ref{sec_chall}) and with perspectives on how current \textit{ab initio} models could evolve to account for the environment in which folding takes place and that ultimately influences the structures active in nature (\ref{sec_future}).

RNA and DNA molecules being chemically very similar at a coarse-grained level by neglecting explicitly the OH group, we
 include in this review models and simulation results for DNA also when the DNA model presents insights on how to consider nucleic acids in general.

\section{Atomistic Models} \label{sec_atom}
Atomistic simulations comprise most of the necessary ingredients to describe the interactions ruling the behavior of RNA molecules.
Through well established empirical potentials, one can in principle access the question of folding and of the dynamics and thermodynamics of the molecule.
Typical atomistic force-fields adopt an harmonic description for bond lengths and angles, sinusoidal potentials for dihedral angles,
Lennard-Jones potentials to describe long-range Van der Waals interactions,
and a description of electrostatics  which is dependent upon the treatment of solvent and ions.
Coulomb potentials in periodic boundary conditions are used for explicit solvent and ions representation, and Born or generalized Born approximations are used in implicit solvent to avoid the computational cost of having to solve Poisson-Boltzmann equation.
For nucleic acids, historically, the reference atomistic force field is AMBER which has been developed and tested thoroughly over the past fifteen years, especially on DNA duplexes \cite{Beveridge2004, Dixit2005}. 
More recently,  the force field parameters have been adjusted to account for A-form helices, the typical helix formed by RNA, 
and to correctly represent short loops, even though results remain in general sensitive to the ionic \cite{Joung2008} and water representation chosen \cite{Hashem2009}.

Even with the most recent force field, ff12, simulating single stranded RNA remains an open challenge for at least two reasons.

The first, obvious, problem is in the size of the systems that can be considered and that limits the applicability of atomistic simulations to small fragments of less than a dozen nucleotides. 
In explicit solvent, the effects of size for nucleic acids can have an even higher impact than for proteins given that RNA and DNA tend to have more elongated configurations, and therefore require larger water boxes, slowing down the simulation even further.
In practice, atomistic simulations on large systems are limited to an investigation around the experimentally available structure.
To this day, only a few unbiased simulations have been able to describe folding events for molecules no longer than a dozen nucleotides.

One recent interesting atomistic simulation highlighting the challenges of single stranded nucleic acid folding has been presented by J. Sponer and collaborators \cite{Stadlbauer2013}.
They study the late stages of folding of DNA G-quadruplexes, a motif that can be formed by both RNA and DNA, found on telomeres and thought to be related to the development of certain cancers \cite{Neidle2010}.
These motifs clearly show the peculiarity of the structures that single stranded DNA or RNA can form, greatly departing from the double helix, in which several bases can interconnect forming hydrogen bonds on all of their three sides (Watson-Crick, Hoogsteen and Sugar) giving rise to "platforms" of three or four bases.
\begin{figure}[tb]
\begin{center}
\includegraphics[width=0.4\textwidth]{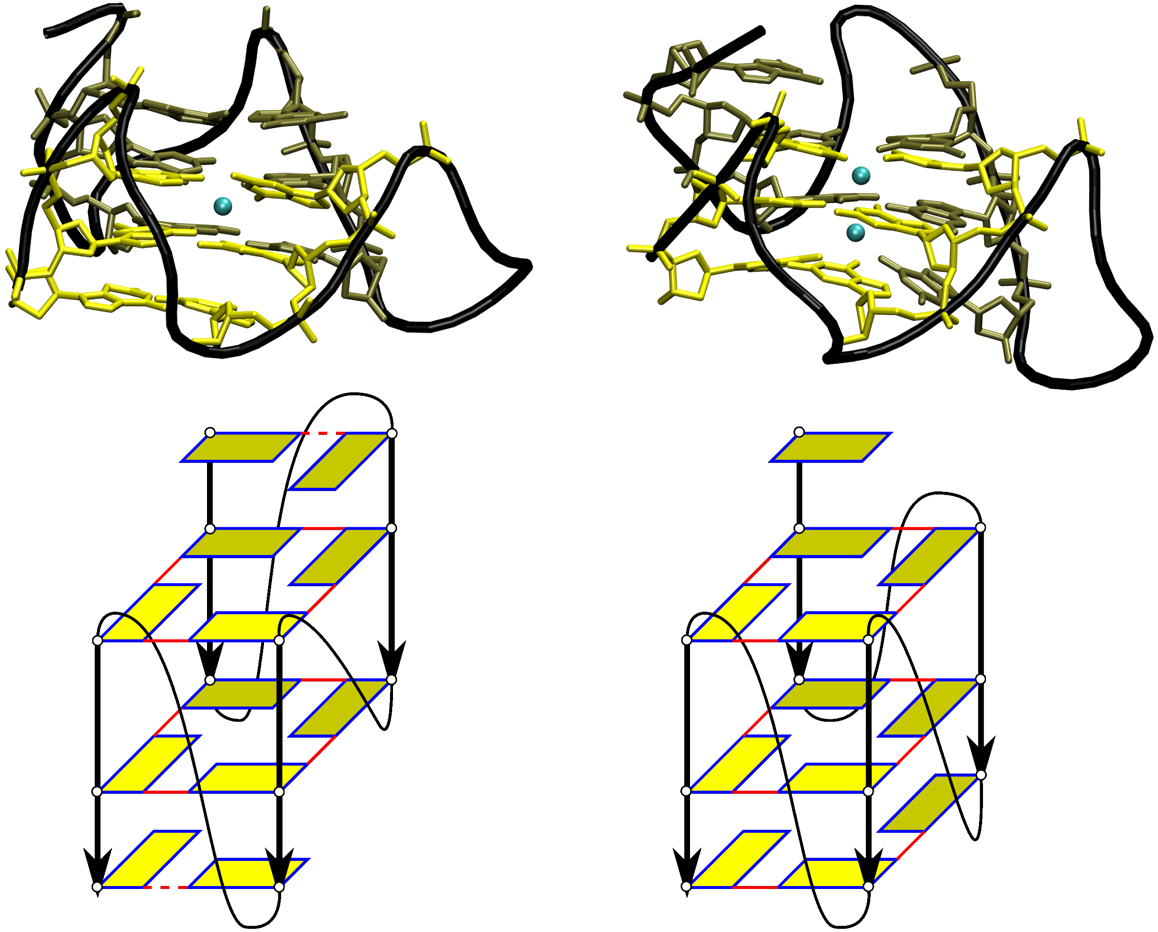}
\caption{{\small Non-native stable G-quadruplexes structures obtained by J. Sponer's group in MD atomistic simulations after partial unfolding by ion depletion (top) and their schematic organization (bottom). }}\label{fig_quad}
\end{center}
\end{figure}
In the case of G-quadruplexes four G bases come together on a plane forming squares through Watson-Crick and Hoogsteen pairings;
several such G-quartet configurations stack on top of each other with intercalating K$^+$ or Na$^+$ ions.
In the simulation by Sponer, the native G-quadruplexes configurations are destabilized by an initial simulation in the absence of the ions.
Under these conditions the quadruplex opens to partially unfolded states
that are then used as initial configurations to attempt refolding once natural ionic conditions have been restored.
Refolding is studied through Molecular Dynamic (MD) simulations at ambient temperature and physiological ionic conditions (figure \ref{fig_quad}).
It is found that not all the partially unfolded configurations are able to fold back to the experimental quadruplet, but some remain trapped in states with alternative base-pairing organizations.
Once alternative base pairs are formed in MD they are too stable to break under physiological conditions, and for all practical purposes the molecule can remain trapped indefinitely, making it impossible to better explore its configurational landscape.

The second, more subtle, problem is related to parametrization.
Most of the structures currently known for nucleic acids come from double helices, implying that if we can characterize well all degrees of freedom of the systems for configurations
close to those adopted by bases forming Watson-Crick pairs, we are left with little information on all other possible structures.
Biases in parametrization procedures are common in all fields, but that is especially problematic in the context of RNA folding where we want to investigate configurations that are heavily under-represented in the databases used to train the force-field.
Indeed, in single stranded RNA many bases are involved in interactions other than those of a regular double helix
and even the sugar-phosphate  backbone can be found highly deformed from the typical helix. 
When structures depart significantly from the double helix, one is no longer guaranteed that the parameter set, or even the functional
forms used, are still adequate to describe the system.

The recent work of Garcia and collaborators on RNA tetraloop folding clearly illustrate these difficulties.
Through REMD atomistic simulations, Garcia was able to obtain full folding trajectories and thermodynamical information for the folding of three 8 nucleotides hairpins forming hyperstable tetraloops \cite{Chen2013}. 
Garcia was able to reach such significant result through an important reparametrization of the AMBER-99 force field to obtain better agreement with thermodynamic and kinetic measurements
of RNA monomers and dimers. 
To derive accurate RNA parameters, the Lennard-Jones interaction had to be fully revisited to correct for AMBER overestimate of base-stacking propensities, imbalance between syn and anti glycosidic rotamers,
and violation of contact distances as calculated by quantum chemical means.
The new parametrization was obtained through an optimization which included many different experimental sources other than structures of the NDB, including in particular thermodynamic data.
Prior attempts to fold hyperstable tetraloops with atomistic details using various folding techniques (fragment reconstruction \cite{Das2010}, interactive simulations \cite{Sorin2005}, REMD \cite{Kuhrova2013})  
failed to predict non-canonical interactions forming in the loops and responsible for its high stability.
The work by Garcia shows that commonly used force-fields for nucleic acids are still far from being optimal,
and as a consequence, even if one was able to simulate larger systems using massive computing resources, 
results might not be reliable, at least for the time being. 

\section{Bioinformatic models} \label{sec_bioinfo}
In order to predict structures for systems of the size of whole RNA molecules one needs to leave the atomistic
description and resort to techniques adopting a simplified vision of the molecule or use bioinformatic methods exploiting existing structural experimental data.
In this section we will focus on approaches base on bioinformatics, while the we will discuss \textit{ab initio} models in the next.

The bioinformatics, or “hybrid”, category comprises a large variety of methods, going from fragment reconstructions, to models strongly relying on secondary 
structure prediction algorithms and 3D scaffolds extracted from the NDB. 
These methods are useful to obtain three-dimensional structures, but provide no information on the dynamics and thermodynamics of the molecule.
\\
Most hybrid models employ a simplified, coarse-grained representation of the molecule, the choice of which depends on the information that is being exploited 
to make the prediction and on the calculations that follow, experimental structural information extracted from the NDB as a structural library, and, often, 
secondary structure prediction algorithms.

Given that bioinformatic methods have been the subject of numerous publications and reviews \cite{Rother2011, Cruz2012, Laing2010, Laing2011, Cao2012,Shi2014}, we will not give here an exhaustive account of the field, but only an illustration of what can be achieved integrating various sources of information from bioinformatics and modeling, later to be compared with the performance of \textit{ab initio} models.

\subsubsection*{Secondary structure prediction}
Before entering the details of bioinformatic approaches, it is useful to spend a few words on secondary structure predictions, as most methods, to varying extent, assume hierarchical folding and base their three-dimensional prediction on the prediction of the secondary structure first. 
The strategy to determine secondary structures consists of looking for the most
stable set of base pairing under a simplified free energy scheme. 
2D free energies are based on sequence complementarity and constructed under the assumption that the energy is additive.
While early algorithms considered only the energies of single base pairs, current models
also include the effect of neighboring base pairs on the free energy, as well as loop lengths
and composition.
The more recent 2D energy models therefore require a large number of parameters,
mostly obtained from calorimetry experiments.
\\
As a first approximation it is assumed that all base pairs are nested. 
In this case the most stable secondary structure can be
found efficiently using dynamic programming algorithms, with a scaling \(\mathcal{O}(N^2)\), with N the sequence length.
A free energy minimization approach including also the evaluation of all possible 2D structures,
and not just the most stable, and taking into account their contribution to the partition function, has a scaling of \(\mathcal{O}(N^3)\).
The RNAstructure server \cite{Reuter2010} and ViennaRNA package \cite{Lorenz2011} both implement the latter approach.
Pseudoknots however are missing from these prediction methods as they are non-nested structures.
While approaches similar to dynamic programming can be applied to pseudoknots, they tend to scale to a higher power of N,
from \(\mathcal{O}(N^4)\) to \(\mathcal{O}(N^6)\), depending on the class of pseudoknots considered \cite{Mathews2006},
severely limiting their application to more complex RNA.

\subsection{Hierarchical predictions}
Several methods use predictions of the secondary structure as starting point to determine the 3D configuration.
Under the assumption that local contacts form first and that helices are more stable than tertiary interactions, the knowledge of the secondary structure allows to greatly reduce the conformational space to be explored in 3D and provides a useful information to build a starting configuration then to be refined.
The predictive power of these approaches depends directly on the accuracy of the secondary prediction.
Most secondary structure prediction methods capable of considering sequences of the size of whole RNAs are based on nested algorithms and account for Watson-Crick pairs only. 
These methods do not provide reliable information for 3D predictions as they miss to represent both pseudoknots and non-canonical tertiary base pairs that are essential for large RNA architectures.

One exception is given by Vfold \cite{Cao2011}, which allows for the computation of free energies for secondary structures including pseudoknots. 
This method is based on a coarse-grained representation of the system that allows a direct evaluation of entropy parameters for different RNA motifs. 
Energies of base-stacking are taken from the Turner energy set \cite{Serra1994} including non-canonical pairs in loops, and entropies are estimated by 3-body virtual bond model \cite{Liu2010}.
In a second stage a 3D coarse-grained scaffold is constructed based on the secondary structure prediction.
Helices are modeled by A-forms and loops and junctions are constructed from fragments from the NDB.
An optimization procedure selects the most stable scaffold, after which a fully atomistic model is reconstructed and refined using AMBER energy 
minimization.

Based on secondary structure predictions, Vfold also makes predictions on melting temperatures and folding intermediates. 
Currently among the best prediction techniques, it was recently made available to the public through an online server \cite{Xu2014}.

\subsection{Fragment assembly}
Fragment-based approaches have been developed for RNA \cite{Das2007, Parisien2008} . 
They use experimentally determined structures
to construct a repertoire of smaller elements, or fragments, associated to short sequences.
For a given longer sequence, fragments are combined,
and the total energy of the system is calculated according to an underlying physical model, which can be atomistic or coarse-grained.
The structure obtained can then be refined using a local optimization or a short Monte Carlo run.
Given one short sequence is usually associated to a variety of possible fragments, many structures are generated and ranked in energy.
\\
These methods have shown a good ability to reconstruct local structures with high precision,
but are less accurate on bigger structures \cite{Das2010}.
In particular, the prediction capability of these methods is limited for structures displaying unusual or unknown folds, as they rely on structural repertoires in databases.

FARNA/FARFAR by Das and Baker \cite{Das2007} is the RNA fragment reconstruction method gemmed from the ROSETTA method \cite{Rohl2004}, so successful for proteins.
At the heart of the procedure is a stepwise assembly of fragments composed of 3 nucleotides, treated in atomistic detail,
to generate several million possible conformation for each given sequence \cite{Sripakdeevong2011}. Though more expensive than usual fragment reconstruction,
this method allows to sample new combinations of nucleotide conformations.
The physical potentials implemented in ROSETTA are used to compute the energies and to rank these configurations. 
Energies can be computed atomistically (FARFAR) or using a simplified coarse-grained potential (FARNA). 
Both energy models can form non-canonical pairs listed in the Leontis annotation, but are limited in the size of the molecule they can study, making it unfeasible to predict long-range tertiary contacts that stabilize large RNA molecules. 
FARNA/FARFAR was able to successfully predict the structure of molecules of less than 40 nucleotides, correctly reproducing the local backbone deformations induced by non-canonical pairings.
The full-atom energy function can be supplemented with harmonic restraints to impose base pairs, typically obtained from secondary structure predictions.

Another example of fragment-based prediction method is the MC-Fold and MC-Sym pipeline, constructing three-dimensional models from a library of nucleotide cyclic motifs that incorporate all base pairs \cite{Parisien2008}.
Adjacent cyclic motifs share common base pairs and allow to propose a secondary structure inclusive of all possible pairings from a fragment reconstruction method.
From a sequence, the MC-Fold method generates an ensemble of 2D structures
ranked by their probability of occurrence, which is estimated based on
the observed probability of the various 2D motifs in the proposed structure, given the sequence.
MC-Sym follows a similar approach, but using 3D motifs,
which are then combined via a Monte Carlo method to generate 3D structures.
This pipeline has shown good accuracy for large structures, and is totally automated.
It was able to build the 3D structure of a precursor microRNA and of a frame-shifting segment of HIV.

\section{\textit{Ab initio} coarse-grained models} \label{sec_CG}
In order to address the broader question of how a molecule attains its fold, of the different folding pathways, of the response of the system to environmental conditions, and of thermodynamic properties, a physical description of the systems is necessary.
In Section \ref{sec_atom} we discussed how atomistic simulations would in principle provide access to all this information, but in practice they are severely limited in the size of the structures that can be studied in attainable times, making it impossible to address the question of folding of full RNA molecules.
An alternative strategy is to give a simplified representation of the system, focusing only on the degrees of freedom that are thought to be relevant to the folding problem.
This can be done either by keeping the atomistic description but freezing some degrees of freedom into rigid bodies, a strategy which has proven useful to address the question of DNA interconversion
between B-form and A-form helices \cite{Mazur2003} but that to our knowledge has not been extensively investigated for RNA single stranded folding, or by adopting a coarse-grained approach, where groups of atoms are replaced by beads with averaged interactions.
\textit{Ab initio} coarse-grained methods try to capture the physics of the system in an effective theory suited for the spatial and temporal scales involved in folding.
Through simulations responding to physical laws, they aim at predicting equilibrium structures as well as folding intermediates, and at investigating the molecule's energy landscape and thermodynamics.

Over the last few years several coarse-grained models have been proposed to address RNA folding, with different level of resolution and different complexity of the force-field.
We can make a first classification of these models based on the level of resolution adopted to represent a nucleotide.
\\
Maciejczyk and Sheraga developed NARES-2P \cite{He2013a}, a 2-particle minimal nucleic acids representation for both DNA and RNA, and showed that dipole interactions between bases are sufficient to drive the  formation of double helices from unpaired single strands, with a potential that is entirely physics based, and not specifically designed to reproduce neither nucleic acids structures nor thermodynamical properties.
\\
Hyeon and Thirumalai developed the Three Interaction Site (TIS) model \cite{Hyeon2005}, a Go-like model with a specific stabilization term for tetraloops.
The model has been used to study mechanical unfolding of hairpins and the stability of some pseudoknots, observing in particular the dependence of folding pathways and stability upon minor sequence variations in molecules with the same topology \cite{Cho2009}.
\\
Dokholyan's group developed iFoldRNA, a 3-beads representation coupled to Discrete Molecular Dynamics (see section \ref{sec_sim}), an enhanced sampling technique giving access to the vast RNA conformational space. 
The model has been extensively tested on over 150 molecules of sizes ranging from a dozen to one hundred nucleotides and was used in the investigation of folding pathways to address the question of folding hierarchy.
\\
Plotkin's group developed a 3-particle DNA model where the beads representing the sugar and the phosphate groups are considered as spherical particles, and bases are treated as ellipsoids \cite{Morriss-Andrews2010}. 
The model was shown to correctly predict persistence lengths of both single stranded and double stranded DNA, and it has been used to study temperature dependence of twisting and stacking of double helices.
\\
Doye's group recently developed a rigid model with 5 interaction sites for both DNA and RNA, optimized on thermodynamic properties \cite{Sulc2014a}. 
The model is shown to be suited for the study of folding of a small pseudoknot, of melting of a kissing complex, of the dynamics of a double-helical nanoring, and of hairpin
unzipping under pulling of the extremities.
The twin DNA model, OxDNA \cite{Sulc2012}, has been successful in the study of large DNA nanostructures and in reproducing results of single molecule pulling experiments \cite{Romano2013}. 
\\
A 5-particle RNA model was introduced by Xia and coworkers. 
In unbiased simulations the model correctly folds several structures of less than 30 nucleotides, including hairpins, duplexes and pseudoknots \cite{Xia2010}, and, when coupled to a limited number of base-pairs restraints and experimental data such as those coming from Small Angle X-ray scattering (SAXS) experiments \cite{Xia2013}, is able to fold structures up to about 120 nucleotides.
\\
Lastly, at the resolution of 6 or 7 beads per nucleotide, depending on the base species, we have developed the model HiRE-RNA \cite{Pasquali2010,Cragnolini2013,Cragnolini2014} in 3 successive versions.
While the earlier versions were able to correctly predict folds of simple hairpins and duplexes, the most recent development allows to consider molecules of complex architectures and larges sizes, and fold a a 49 nt triple helix pseudoknot from knowledge of the sequence only \cite{Cragnolini2014} as well as a 80nt riboswitch when three base-pairing constraints are imposed.
\\
In what follows we are going to discuss some key ingredients for building a sensible coarse-grained model, focusing on designing the force-field, on parametrization, and on how to include the interactions specific of nucleic acids and of RNA in particular. 

\subsection{Coarse-grained representation}
The first element going into building a coarse-grained model is the choice of the representation, which, in physics terms, starts with determining what are the relevant degrees of freedom of the system for the process under investigation.
One needs to define the number and type of elements that are going to constitute the new particles, grains or beads,  that are then going to interact via an appropriate force-field.
The choice of the beads reflects the degree of resolution adopted and is directly linked to the ability to reconstruct back an atomistic model from the coarse grained representation.
A detailed model, with many beads, is clearly computationally more costly than a simple model with a few beads, therefore the choice of the representation is a trade off between speed and accuracy and depends sensibly on the questions that are to be addressed by the model.
Models with rigid bodies or with two or three beads per base have proven useful to study duplex assembly and melting processes, but lack many details necessary for structure prediction. 
Models with 5 or more particles, thanks to a more accurate representation of the bases, allowing to define more properly stacking and base-pairing interactions, are better suited for structure prediction, but require more computational time and can lack enough sampling for the study of thermodynamic properties.

\begin{figure}[tb]
  \centering
  \includegraphics[width=0.5\textwidth]{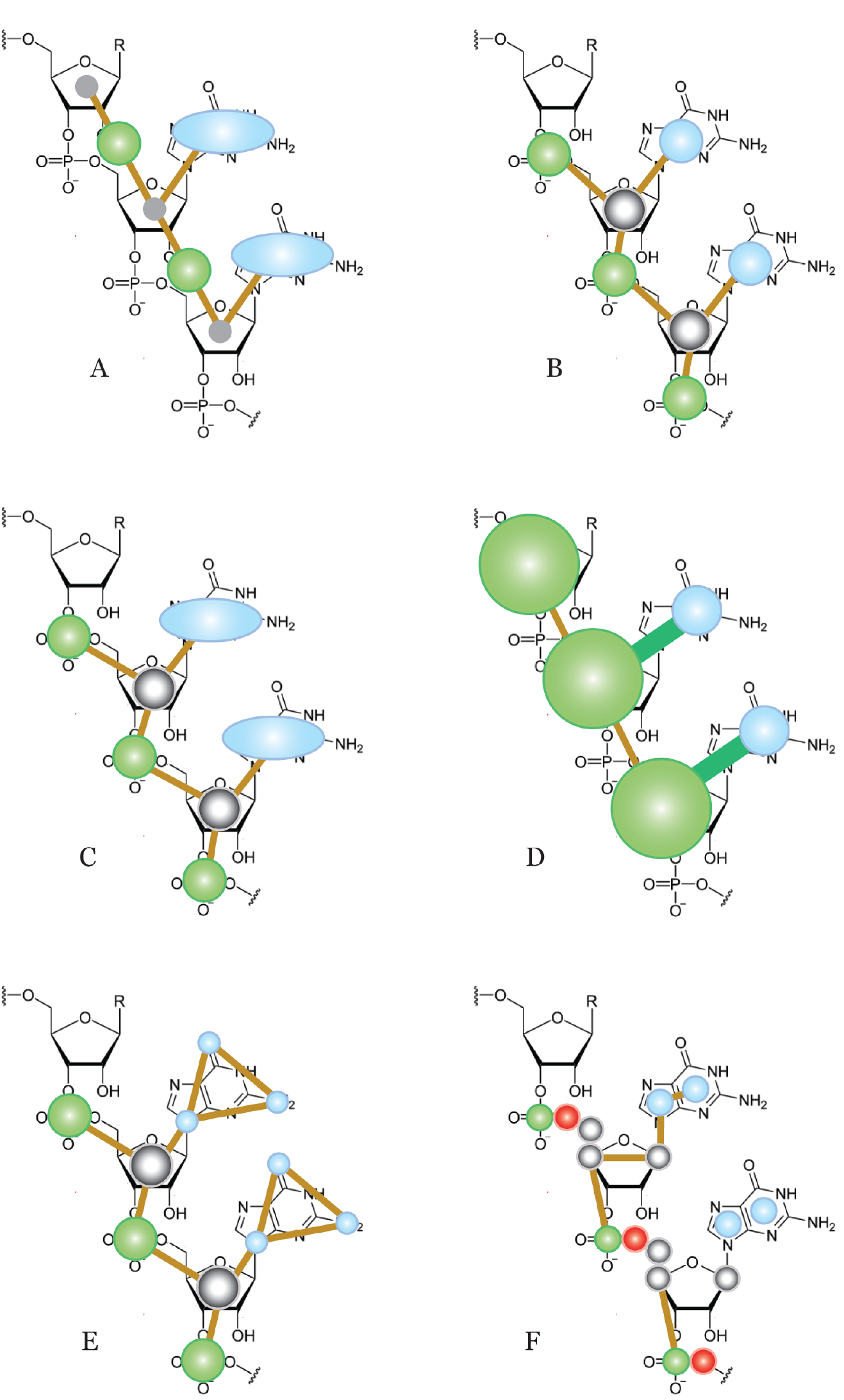}
  \caption{\label{fig_beads} Coarse-grained representation of various models: NARES-2P (A), iFoldRNA and TIS (B), Plotkin's (C), OxRNA (D), Xia's (E), HiRE-RNA (F).
Beads centered at the phosphates are shown in green, beads representing sugars are in grey, and beads representing bases are in blue. 
Lines connecting beads are in brown when beads are connected through flexible potentials and in green for rigid bodies.}
\end{figure}

The models mentioned in the previous section show well the wide range of possible choices.
NARES-2P defines 2 beads, one spherical for the phosphate and one elliptical for the base, plus a virtual sugar used in the definition of the relative geometries of the phosphate and the base, but that does not participate in the interactions.
TIS and iFoldRNA define 3 spherical beads: a phosphate, a sugar and a base, positioned at the center of mass of the respective groups.
Plotkin's model defines 2 spherical beads, representing the phosphate and sugar groups, and one elliptical bead representing the base.
OxRNA defines the nucleotide as a rigid body composed of 5 interaction centers in different locations depending on the kind of interaction considered.
Xia's model defines 5 particles: a phosphate, a sugar and 3 particles for the base positioned differently according to the base type.
HiRE-RNA defines 6 or 7 particles: four in the positions of the backbone's heavy atoms P, O5', C5' and C4', one on the sugar C1', and one or two beads in the center of mass of the aromatic rings of the bases.
In figure \ref{fig_beads} we give a summary of the bead representations of the different models.

The choice of the representation is strictly coupled to the choice of force-field.
For example, it is clear that if we want to include stacking interactions the model needs to have the possibility of defining a plane for the bases through a sufficient number of beads, through ellipsoids, or through an internal reference frame.

\subsection{Up or Down?}
Once the particles of the model have been chosen, one needs to give them properties on how to interact, namely define a force-field though potentials.
The introduction of a force-field is what makes this approach physical, or \textit{ab initio}, as opposed to a bioinformatic, data-mining, approach.
Once the potentials have been defined, the system obeys to classical mechanics, with forces computed as spatial derivatives of potentials, accelerations computed through the inertia law, and particle trajectories obtained through integrations over time.

Among \textit{ab initio} models we can make the distinction between those for which the force-field is built systematically from integration of the underlying degrees of freedom, called 
“bottom-up” models,  and “top-down” models that, with varying extent, make use of experimental data to assign parameters of a potential assigned a priori
or derive statistical potentials all together.

\subsubsection*{Bottom-up}
Bottom-up potentials follow the natural definition of a coarse-grained model, where fast degrees of freedom are integrated over and included into the interactions
of the slower variables. 
Features of these potentials are usually extracted from long atomistic simulations. 
Two examples of bottom-up potentials are NARES-2P and the model by Plotkin.

In NARES-2P the interaction between bases and phosphates is described through four local interaction energies - bond-stretching harmonic potential, an angle bending sinusoidal potential, a torsion sinusoidal potential, and a sugar-base rotameric potential - and non-local interactions between bases and phosphates.
The local energy terms were fitted to the Boltzmann inversion of the respective distributions obtained from the PDB structures of several dozen DNA and RNA molecules though a common structure-based optimization procedure.
Non-local terms include a base-base Gay-Berne potential accounting for close contact repulsion and long-range attraction of nonspherical beads \cite{He2013a},
a dipolar base-base electrostatic interaction, a base-phosphate and phosphate-phosphate Lennard-Jones potential, and a Debye-Huckel electrostatic potential between phosphates.
To describe the anisotropy of the beads representing the bases, the analytical expression of base-base interactions are rather articulate and involve 11 parameters for each one of the allowed 15 base pairs.
Both non-local base-base interactions were parametrized fitting potentials of mean force (PMF) computed by numerical integration of AMBER energy surfaces, through a systematic averaging over the degrees of freedom not represented in the coarse-grained model.
To make the calculation feasible, integration was performed on a grid. 
Potentials derived with this procedure were shown to still represent well the directionality of the potentials computed with AMBER for atomistic structures, i.e. of the potential prior to integration.
Some parameters were left free from the PMF fitting procedure and were then adjusted based on the nearest-neighbor parameters of Santa Lucia's HyTher model \cite{SantaLucia1998} in order to reproduce the thermodynamics of DNA folding.
\\
The interaction between phosphates depends significantly over the environment surrounding the charges as it is screened by water molecules and counter ions in solution.
Phosphate-phosphate interaction energies were derived from an umbrella sampling atomistic AMBER simulation of two phosphate ions in a TIP3P water box and counter ions.
A potential of mean force was then extracted averaging over the degrees of freedom of water molecules and counter-ions.

Another model built from a bottom-up procedure is Plotkin's DNA model.
Its potential is composed of local bond, angle and dihedral terms, electrostatic interaction between phosphates, non-local base-base, base-residue and residue-residue interactions, (where a residue can be either a sugar or a phosphate) accounting for the elliptical shape given to the bases, and base-base hydrogen bonds for base-pairing.
The functional form adopted for Van der Waals base-base interactions is a modification of the Gay-Berne potential called RE$^2$ potential \cite{Everaers2003,Babadi2006} with 14 parameters for the ten possible base-base interactions.
These parameters are determined by fitting RE$^2$ and the all-atom molecular mechanics force field MM3 \cite{Lii1989} with a Buckingham exponential-6 potentials for long-range interactions.
The same procedure is adopted for the optimization of base-sugar or base-phosphate parameters for which the interaction potential is the limiting case of one ellipsoid interacting with a sphere, while the sugar-sugar and sugar-phosphate Van der Waals interaction are described by a Lennard-Jones potential with pre-fixed equilibrium distance and depth, to prevent steric overlap of the backbone particles.
Base-pairing is not included in the RE$^2$ potential and it is described by a phenomenological 12-10 LJ and sinusoidal angular dependence potential, where the angles account for the relative orientations of the two interacting ellipsoids and are defined through their normal vectors.
Geometric parameters parameters are pre-set from the specific configurations of base-pairs and the maximal energy of the pair (bottom of the potential well) is also given \textit{a priori} based on energy calculations \textit{in vacuo} and on experimental evidence of the role of hydration on hydrogen bonds.
Electrostatic interactions between phosphates are described by a Debye-Huckel potential with parameters given to represent a system immersed in water and at a fixed ion concentration (200nM), i.e. fixed screening length.
For all local potentials no functional form is assumed \textit{a priori}, but the forms and parameters of the potential are extracted from equilibrium all atom simulations.
In order to obtain results that are not including effects of other interactions for which the functional form of the potentials are imposed from the start, a modified system with no base-base interactions and minimal Coulombic interactions was designed.
The modified system is simulated for 250ns using CHARMM27 parameter set, in explicit water and with counter ions.
The long simulations are required to ensure a good convergence of the extracted potentials.
It is interesting to notice that potentials derived according to this procedure can adopt significantly different forms from commonly assumed phenomenological potentials between the same set of atoms.
For example of the 11 different bond angles in the coarse-grained model, 5 were fitted by harmonic potentials, while the remaining 6 were better fitted by a double well potential.

As these two example clearly show, bottom-up potentials still require the assumption of the functional form of most interactions, but the coarse-grained parameters are derived through fitting the underlying atomistic potentials and are obtained either directly through integration of the atomistic force-field, or indirectly, through PMF computed from atomistic simulations.
Bottom-up models find their limits on the validity of the underlying atomistic force fields.
As discussed in Section \ref{sec_atom} this could reveal to be a severe limitation as for the time being atomistic potentials, even in their most refined version, 
are known to miss-represent some interactions, such as those occurring between phosphates and bases \cite{Zirbel2009} and the account of ions in solution.

\subsubsection*{Top-down}
Top-down models are effective theories where the phenomenological interactions between the particles of the system is given \textit{a priori} based on physical intuition and 
parameters are obtained through a systematic procedure exploiting different kind of experimental evidence.
Three examples of top-down models are OxRNA, Xia's model and HiRE-RNA, all defining different force fields and optimization strategies.
OxRNA parameters are fitted to thermodynamic data, while both HiRE-RNA and Xia's model are optimized based on structural information of the NDB.

OxRNA force-field is composed of a backbone interaction term, modeled by a finitely-extensible nonlinear elastic potential, an excluded volume term modeled by a Lennard-Jones potential, a stacking term and a base-pairing term both modeled by a Morse potential, with the stacking term including also an explicit linear dependence with the temperature, a coaxial and cross stacking terms both modeled by harmonic potentials.
In the absence of sufficient experimental thermodynamic data for small molecules, predictions made with Turner's nearest-neighbor model (NN-model) were used to derive melting temperatures for a large set of small RNAs containing different motifs.
Fitting of interaction strengths were done by simulated annealing to find a parameter set minimizing the differences between the melting temperatures calculated via the NN-model and those extracted from OxRNA simulations on the same systems.
OxRNA was first optimized to reproduce melting temperatures of structures including only canonical pairs to obtain an averaged parameter set independent of the sequence specificity.
In a second stage sequence specificity was introduced for Watson-Crick and wobble base pairs and optimized to fit melting temperatures of a large set of short sequences forming hairpins or duplexes. 

The Hamiltonian of Xia's model is composed of a set of bonded terms, including bond stretching, angle bending and dihedral energy, and a non-bonded effective potential inclusive of both Van der Waals and electrostatic contributions, modeled through a Buckingham potential \cite{Xia2013}.
Local interaction potentials were derived directly from Boltzmann inversion of variables distributions obtained from 668 3D structures containing more than five base pairs, resulting in the usual harmonic functions for bond lengths and angle bending, and sinusoidal form for dihedrals.
Non-bonded parameters were fitted to reproduce global energy minima and later refined to minimize the difference between energy-minimized coarse-grained structures and their corresponding experimental structures.
The parameter set was then validated through the comparison of coarse-grained simulations and atomistic simulation on a set of 15 molecules.

HiRE-RNA's potential is composed of local harmonic terms for bond angle stretching, sinusoidal energy for dihedrals, excluded volume, Debye-Huckel electrostatic energy,
and specifically designed stacking and base-pairing terms keeping into account base orientations \cite{Cragnolini2014}.
The model has geometric parameters whose values have been determined from distributions extracted from 200 NDB structures including molecules of varying sizes and topologies; overall energetic parameters, representing the relative weights of the different interaction terms, which are subject to an optimization procedure; and base-pairing energetic parameters, which for the time being are assigned from the start based on the number of hydrogen bonds of the contact, and no longer modified.
The optimization procedure is done through a genetic algorithm to find the parameters that better distinguish energetically native structures from decoys \cite{Maupetit2007a}.
For each structure of a training set we generated 20 decoys including low energy and high energy structures. 
Low energy decoys were chosen to evenly cover four possible scenarios of high or low rmsd and high or low base-pairing similarity with respect to the native structure, in the goal of covering extensively the different possible conformations adopted by a given sequence.
The algorithm mimics an evolutionary process in which vectors containing the energetic parameters undergo mutations and swapping to obtain a combination of parameters that maximizes the energy difference between the native structures and all decoys.
To optimize with a genetic algorithm the choice of training set is also important.
Since our goal is to have a model that is able to follow a molecule's large conformational changes, we want to have parameters that allow all possibilities, and that are not biased toward some specific conformations.
In particular, for RNA, the risk is to have parameter sets highly favoring helices, given that they are by far the most common structural element in the NDB.
We therefore used the concepts of RNA graphs to build a structure database rich in different topologies \cite{Gan2003, Pasquali2005} since this descriptor captures well the different overall organization of the molecule's structure.
From the RAG database \cite{Izzo2011}, we have chosen one an equal number of representative structures for each populated topology to be part of our training set.
Parameters obtained with this procedure were then tested through long MD simulations on systems of various size and showed a significant improvement over the previous parameters calibrated by hand.

Top-down models require the optimization of many parameters, a task that, depending on the detail of the force field, can quickly become as challenging as parameterizing an atomistic force field.
These models rely intrinsically on the availability of experimental data with a direct correspondence to quantities that can be extracted from the model, such as melting temperatures and spatial variable distributions.
As we saw for OxRNA, even though in principle melting temperatures are experimentally accessible, in practice for a well grounded optimization one needs information on so many molecules that the only viable route is to randomly generate structures to simulate and use thermodynamic models to compute their melting temperatures. 
\\
Geometric distributions are readily accessible from the NDB, but the choice of the structures used to compute them is critical.
As it is the case for bioinformatic prediction models, the risk is to bias configurations toward the double helix, given the large majority of nucleic acids structures in the NDB are of this form.
\\
Much harder is assigning relative weights to sequence dependent base-pairing and stacking. 
Energetic information on these two terms can only be inferred indirectly from thermodynamic data
and single molecule pulling experiments, where the contribution of different energy terms can't be easily disentangled. 
Base pairing and base stacking energies can in principle computed by quantum mechanics calculations,
but for the time being these data are available only for bases in vacuum and in gas phase \cite{Sponer2002,Sponer2004},
and it is unclear how they transfer to the context of a molecule under physiological conditions.
Assigning base-pairs relative weights is the main difficulty in the parametrization of HiRE-RNA, where bases can interact on their three sides on different positions, for a total of 28 possible different pairings.
Each pair contains one, two or three hydrogen bonds.
The choice we have made for now has been to give a pair a weight proportional to the number of hydrogen bonds formed, but this is clearly in contrast with the observed overwhelming abundance of GC and AU canonical pairs in the NDB and with results of QM calculations. 
Indeed we found that to better account for structures we needed to artificially modify these parameters, giving a slightly higher weight to canonical pairs over all others.
From a physical stand point, such difference could derive from a cooperativity effect of hydrogen bonds.
It is to be noted that this problem affects atomistic models as well and therefore bottom-up models are subject to the same uncertainties.

As it emerges from this brief discussion, parametrization is possibly the main challenge in the development of a force field.
Recently, methods have been proposed to consistently integrate both experimental and theoretical data in an automatic parameter optimization procedure\cite{Wang2014}, and they have been applied as a test case to the parametrization of different water models.
So far, these methods are highly expensive, and their efficiency have yet to be demonstrated on more complex models, such as the ones used for protein and RNA.

\subsection{Flat or Round?}
The specificity of nucleic acids interactions in folding is given by base stacking and hydrogen bonding to give base pairs.
Both interactions are dependent upon the flat shape of the aromatic rings of the base.
Typical coarse-grained models of biomolecules developed in the past for proteins and for lipids represent the newly defined particles (grains) as isotropic spheres.
It is the case of the popular Martini force field \cite{Marrink2013}, and of the protein model OPEP \cite{Sterpone2014}, that we also develop, for which both the backbone heavy atoms and side chains are represented by spheres of a appropriate size.
Such an approach find its reasons in a model that is at low resolution, for which interactions are taken as isotropic and long-ranged, often modeled by Lennard-Jones potentials. 
The first coarse-grained models developed for RNA also adopted this description.
For example, both iFoldRNA and TIS describe a nucleotide as composed of three spherical beads.
The specificity of base-pairing and of helix formation are integrated into iFoldRNA by giving a set of distance constraints to the base beads, including both same strand and on the cross strand terms.
In TIS it is the Go-like potential that drives the molecule to the native base-pairing and stacking.
In neither of these models bases are really free in their interactions, in one model because they are constrained by geometry, in the other because they are biased toward the native structure.

If we want to model base behavior realistically, we need to look closer at the base, adopting a relatively high resolution that takes into account the anisotropy of stacking and hydrogen bonding.
The more recent nucleic acids coarse-grained models adopt different strategies to take into account base planarity and orientation, going from an ellipsoid base representation (Plotkin and NARES-2P), 
to introducing an internal reference frame and several interaction sites (OxRNA), or to explicitly define planes thanks to the high resolution description (HiRE-RNA). 
\begin{figure}[tb]
  \centering
  \includegraphics[width=0.45\textwidth]{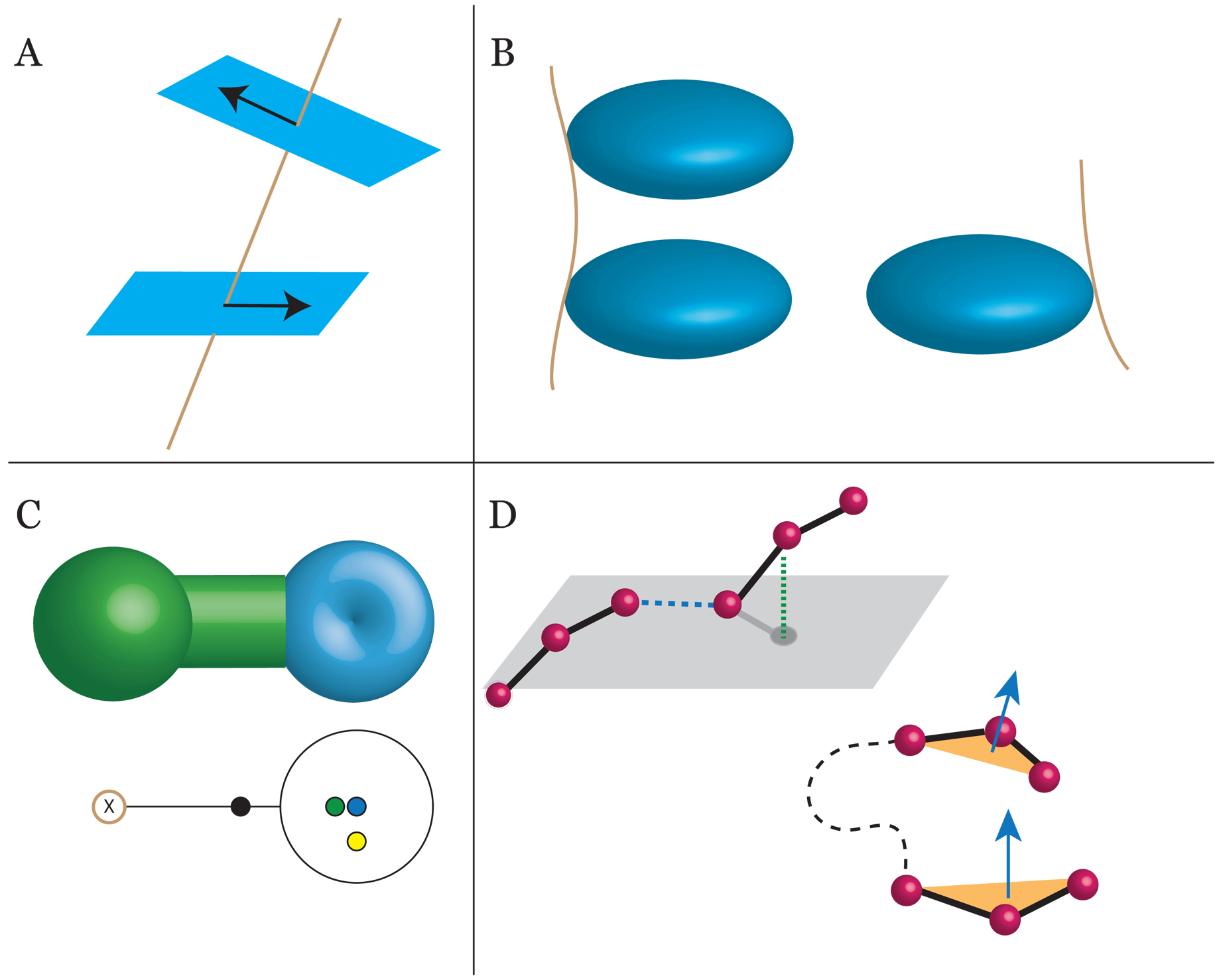}
  \caption{\label{fig_flat} Top left: Plotkin's model base representation through ellipsoids, top right: OxRNA base representation through several interaction centers, bottom left: HiRE-RNA base representation through three beads and normal vectors, bottom right: NARES-2P components of the bases dipole moments entering the base-base dipolar interaction.}
\end{figure}

In NARES-2P bases are subject to an electrostatic dipole-dipole interaction and a non-bonded interaction, including excluded volume and long-range Van der Waals.
Both are modeled through anisotropic potentials, with the dipole interaction dependent on the angle of the bases
with respect to the distance vector (figure~\ref{fig_flat} A) and the long-range potential modeled through a Gay-Berne potential
under which bases perceive each other as ellipsoids. 
The interaction between bases and phosphates is on the other hand modeled through a standard LJ potential, as if the bases were spherical.
In the model by Plotkin, bases are always subject to a modified Gay-Berne potential, interacting with other bases as two ellipsoids for stacking and pairing, and an ellipsoid-sphere couple when interacting with phosphates or sugars (figure~\ref{fig_flat} B).
As on one hand, representing bases as ellipsoids seems the natural choice, the expressions going into defining elliptical potentials are non trivial and computations more demanding than when dealing with spheres.
This choice of description is also not practical when one wants to introduce the possibility of base-pairing on the different sides of the base, as it would be necessary to introduce an angular inhomogeneity of the ellipsoid to distinguish Watson-Crick, Hoogsteen and Sugar sides, and to modify accordingly the already articulate functional form of the potential.

OxRNA considers nucleotides as rigid bodies and defines isotropic potentials for the interaction of the bases.
However, in contrast to simpler models, thanks to an internal reference frame, the model defines different interaction centers for different kinds of potentials, effectively breaking the spherical symmetry.
Of the 5 interactions centers, 4 determine the base behavior with respect to hydrogen bonding, coaxial stacking, 3' stacking and 5' stacking (figure~\ref{fig_flat} C). 
This method has the advantage that all potentials assume simple functional form, either Morse, Lennard-Jones or harmonic, modulated by a smoothing function bringing them to zero at long-distances, making calculation practical and economical.
On the other hand, with the use of a rigid body description important structural details can be lost.
It is important to keep in mind, though, that the purpose of the model was never to give an accurate structural description, but to obtain the thermodynamical behavior of the molecule, and in this respect the details of rearrangements internal to the nucleotide or the base, are mainly irrelevant.

In HiRE-RNA we take advantage of the high resolution of the base representation to define base planes and we construct potentials using the norm vectors perpendicular to the plane.
Base-pairing potential is composed of the product of a hydrogen bonding potential, depending on the distance and relative angles of the interacting particles (the base extremities),
and a planarity term, where co-planarity of the bases is implemented through a short range inverted Gaussian potential dependent on the distance of the particles of one base with respect to the plane defined by the other base (figure~\ref{fig_flat} D).

Stacking is also dependent upon norm vectors both for base orientation, with a preferred parallel orientation for stacked bases, and to ensure that stacked bases are coaxial.
Our method has the advantage of keeping a spherical particle description and to allow to easily make the distinctions of the different sides of the base and define sideways base-pairs, that, as we will discuss in the next section allows us to define non-canonical and multiple base pairs. 
The drawbacks are in the need of a finer description, and therefore less computational speed since we need 3 beads for each base to properly define a plane, and in the functional forms of the potentials that become multi-body, with the planarity term requiring the contribution of 6 different beads. 
Moreover, there aren't any standard potentials to describe the interactions we want to model, and the functions we have introduced are very much empirical.

\subsection{Watson-Crick or Non-canonical ?}
A detailed analysis of RNA structures has shown that there exists of the order of one hundred possible base-pairings between RNA bases 
since bases have in principle the ability of forming hydrogen bonds on all their different sides \cite{Lemieux2002}.
Following the classification of bases sides introduced by Westhof and Leontis as Watson-Crick (WC), Hoogsteen (H), and Sugar (S), interactions between bases are found to involve all sides combinations.
WC-WC base pairs are the most common, respecting the canonical DNA pairing scheme A $\cdot$ U, G $\cdot$ C, but all other pairings of all sides with each other and of all bases with each other are also found.
Even adopting a simplified view where only one possible pair is formed on each side of the base, considering all 9 possible side-side pairs (WC-WC, WC-H, WC-S, H-H, H-WC, H-S, S-S, S-WC, S-H) in the cys and trans conformation, for all the 12 possible combinations of base kind (GG, GA, GC, GU, AA, AG, AC, AU, CC, CG, CA, CU, UU, UG, UA, UC), we can count over 200 different possible base-pairs.
These non-canonical interactions are especially relevant for single stranded molecules that don't have an WC complementary strand immediately accessible, and are therefore specific to RNA (and to ssDNA).

Only a few prediction models take into account the possibility of forming non-canonical pairs.
As we have seen in Section \ref{sec_bioinfo}, MC-Fold and FARNA include the possibility of forming non-canonical pairs, but 
of the \textit{ab initio} coarse-grained models, most lack the level of details necessary to describe the base sides.

\begin{figure}[tb]
  \centering
  \includegraphics[width=0.48\textwidth]{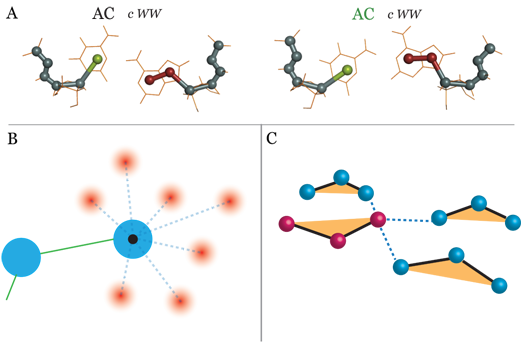}
  \caption{\label{fig_BP} A: Two distinct AC base pairs included in HiRE-RNA, both occurring as cys WC-WC, according to Leontis' classification, and occurring in the NDB with similar probability among the most frequent AC base pairs. B: Different possible interaction centers for the interaction between two bases in HiRE-RNA. C: schematic formation of a base quadruplet in HiRE-RNA. All interactions occurring in natural RNA G-quadruplexes are included in our model.}
\end{figure}

Only the model by Xia and HiRE-RNA can make the distinction between interactions occurring on different sides of the base.
In the model by Xia 3 particles are used to define a base, forming a triangle, with each bead corresponding to a side of the base.
The model includes 14 possible nominal pairings, corresponding to all possible base-pairs for the 4 bases and 3 sides, without considering trans and cys conformations, with average distance parameters. 
Base-pairing is described as part of a generic long-range interaction between beads and depends only on the relative distance between the particles. 
Constructing pairs based solely on the information of the side on which they occur, implies averaging over many possible different interactions occurring between the two sides, which can vary significantly in equilibrium distances, angles and strengths, depending on how many hydrogen bonds are formed simultaneously.
With a more detailed approach, HiRE-RNA, in its current version, includes 28 different possible interactions occurring on all sides, 
each associated to a specific set of distance, angles, torsions and number of hydrogen bonds formed. 
The choice of 28 interactions is quite arbitrary and can be extended to any number of interactions as long as they are sufficiently distinct in interaction centers.
For now pairs have been chosen based on their abundance in the NDB, making sure to have at least two or three representative for each letter pair.
For some letter pairs we can account for two distinct interaction sites occurring between the same sides at different geometric centers (figure \ref{fig_BP}A).

The potential energy is given by a narrow inverted Gaussian around the geometric center, and for any given letter pair, we simply add over all possible centers (figure \ref{fig_BP}B).
Because of the excluded volumes of the beads, effectively, there can only be three interaction centers simultaneously present around a base, one on each side (figure \ref{fig_BP}C).
Despite the fact that HiRE-RNA considers at the moment fewer possible interactions than Xia's model,  coupling hydrogen-bonding with planarity allows to capture fine structural details of base-pairing that can then have a large repercussion on the overall conformation of the molecule, as it is for the formation of triplets and quadruplets.

\section{Simulation Methods} \label{sec_sim}
\textit{Ab initio} models constitute a physical description of the system.
Their natural application is for Molecular Dynamic simulations (MD), where the system is subject to Newton's equations.
If the goal is to observe phenomena occurring on long time scales such as folding, for molecule of the size of most RNAs, simple MD at a fixed temperature
is often not sufficient even with the sensible reduction in degrees of freedom of CG models.
In addition to MD, enhanced sampling techniques are commonly used.
In this section we are going to review the most prominent enhanced sampling techniques that have been used with the various CG RNA models, with particular attention to 
Parallel Tempering (Replica Exchange MD - REMD)  and Simulated Tempering (ST) \cite{Nguyen2013}, both employed by HiRE-RNA, Discrete Molecular Dynamic technique, employed by iFoldRNA, and interactive simulations, an innovative technique, that we are currently testing in combination with HiRE-RNA.
Other simulation techniques, such as Monte Carlo and simulated annealing, are also commonly used for structure predictions, but we will limit here the discussion to MD enhanced sampling variants, more naturally linked to the folding process. 

\subsection{Parallel tempering / replica exchange}
In parallel tempering molecular dynamics (PTMD), a number of MD simulations are run concurrently, with different values of a control parameter.
During the simulation, exchanges of configurations between the replicas are attempted, according to a chosen protocol.

\begin{figure}[tb]
  \centering
    \includegraphics[width=0.45\textwidth]{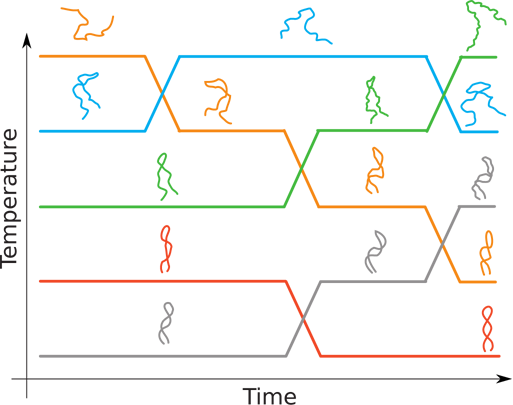}
    \caption{\label{fig_remd} Illustration of a temperature REMD run: at low temperatures the molecule's behavior is led by its internal interactions and it adopts a folded state. At high temperatures the dominant factor is the kinetic energy and the molecule unfolds. The exchange between high-temperature and low-temperature replicas allows the molecules to escape energy minima and explore significantly different conformations.}
\end{figure}

The most common protocol, Temperature Replica Exchange Molecular Dynamics (T-REMD), uses replicas simulated at different temperatures.
The attempted exchange between neighboring replicas (figure \ref{fig_remd}) is done at fixed time intervals and  must obey the balance condition in order for each replica to sample the correct ensemble. 
The most common method is to attempt an exchange between neighboring replica considering energy differences and a metropolis criterion.
This produces a canonical distribution at each temperature, and coupled with a reweighting method such as WHAM\cite{Kumar1992},
allows to recover thermodynamic information about the system, for example its heat capacity.
Though this method will produce the correct ensemble given enough simulation time, the simulation's convergence to the equilibrium distribution can be made faster by repeating the exchange moves a number of times at each exchange pause \cite{Chodera2011} .
The repetition of exchange attempts represents a negligible computational cost, while potentially providing large gain
in the efficiency of the simulation's sampling.

The difference between temperatures (or the chosen parameter) greatly impact the frequency of successful exchanges, and the convergence of the simulation.
The exchange rate will be proportional to the overlap of the chosen parameter's distribution between the replica.
Since the distribution for temperature becomes narrower as the number of particles increases, the number of replica required to cover a given range of temperature increases as larger systems are studied \cite{Earl2005}.
This  problem can be alleviated using coarse-grained systems or by switching to alternate general ensemble methods, such as simulated tempering.

\subsection{Simulated tempering}
\begin{figure}[tb]
  \centering
  \includegraphics[width=0.45\textwidth]{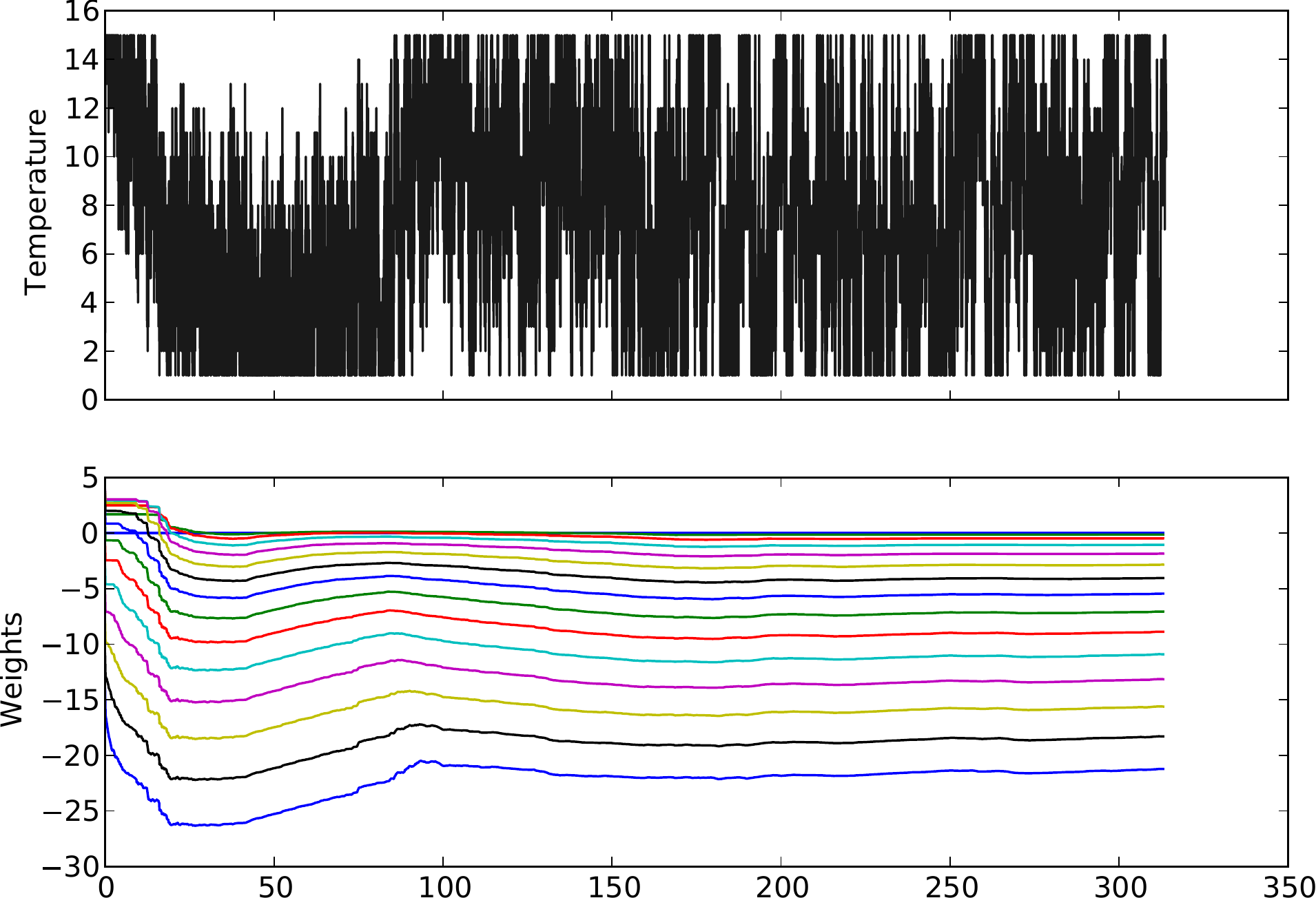}
  \caption{\label{1F9L_ST} Evolution of the random walk in temperature during a ST simulation with on-the-fly update of weights.
  After an initial equilibration phase, the weight factors proportional to the Helmholtz free energies converge and the simulation performs a uniform walk in temperature.}
\end{figure}

Simulated Tempering (ST) is a simulation technique that enhances sampling by raising and lowering the temperature sequentially in time.
The temperature becomes a dynamical variable, taking values in a discrete range $T_1 < T_2 < \cdots < T_N$.
The exchange between temperatures is governed by weights that need to be assigned at the beginning of the simulation to ensure a uniform random walk in temperature space.
With the correct weights ST has a higher acceptance ratio than PT \cite{Mitsutake2000,Park2008}, however, given the weights depend on the Helmholtz free energies at each temperature \cite{Chelli2010}, determining them \textit{a priori} is problematic.
Recently introduced on-the-fly weights determinations, however, allow to obtain the weights automatically,
greatly simplifying the use of ST \cite{Nguyen2013} (an example is provided in figure~\ref{1F9L_ST}).

Each ST simulation being independent, it can easily be parallelized at little additional cost,
making it ideal to run on a large number of CPUs, allowing for faster data gathering. 
ST can also be readily generalized to a random walk with several parameters \cite{Mitsutake2009},
without requiring more than a single simulation, while a similar PT simulation would require
an exponentially larger number of replicas.

\subsection{Discrete molecular dynamics}
Discrete molecular dynamics uses a simplified representation of the energy function, replacing it by discrete step functions.

Instead of using the derivative of the energy function to integrate Newton's equation of motion,
a collision detection algorithm based on the ballistic motions of the particles is adopted.
Atoms move freely until they collide, and since collisions are purely local, only nearest neighbors
need to be considered for possible collisions. 
Since only the particles involved in a collision need to be considered,
fewer updates are required \cite{Proctor2011}.

This method largely decreases the computational cost compared with MD:
the potential functions are much simpler, thus less costly to calculate, and derivatives are not needed.
In traditional MD, the evaluation of energy and force derivatives is the main computational bottleneck.
The main drawback of DMD is the modification of the potential function resulting in an altered kinetics of the system, making it harder, though not impossible, to extract kinetic properties \cite{Buchete2008}.
Though better approximations of the original function
can be constructed by reducing the step size used when discretizing the potential, this result in an increase in computational
cost, until it reaches that of classical MD.

\subsection{Interactive simulations}
Interactive simulations applied to macromolecular manipulation is now an active field of research \cite{Lv2013}.
One recent application has been for fitting models into experimentally determined envelopes \cite{Birmanns2011,Molza2014}.

Interactive simulations are built on the idea of rendering accurate molecular models real and tangible for scientist.
The modest technical requirements allow to set up an interactive simulation session on a small laptop computer, simply controlled by a touchpad or a mouse.
By including the possibility of interacting directly with the simulation,
it is possible to model structural changes in a very intuitive way, and to probe the stability
of structures by directly perturbing the structure \cite{Surles1994}.
This type of approach has seen great success with the emergence
of game softwares challenging players to fold structures
by hand.
With FoldIt\cite{Cooper2010} players showed great performance in predicting protein folds,
and were able to solve a novel protein structure\cite{Khatib2011a}.
The strategies used by the top players were shown to outperform the best prediction algorithms
published so far\cite{Khatib2011}, and the success of FoldIt inspired other similar projects,
notably the EteRNA game for 2D RNA structure prediction\cite{Lee2014}.

The coarse-grained representation is the natural partner for virtual interactive experiments as it represents an excellent compromise between simulation speed and biological fidelity.
Moreover, CG models are in general more robust with respect to user interactions than computations carried out at an atomistic level. 

\begin{figure}[tb]
  \centering
  \includegraphics[width=0.45\textwidth]{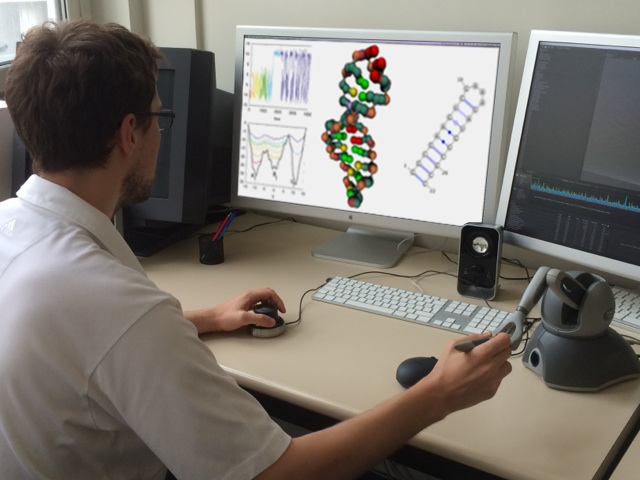}
  \caption{\label{fig_interactif} Interactive simulation of an RNA hairpin with HiRE-RNA interfaced to UnityMol.
  A force is applied from the mouse of from the haptic arm (bottom right) to a particle of the molecule to modify its structure, and can then be released to explore a new configuration. Both the instantaneous internal energy and secondary structure can be followed in their evolution over time.}
\end{figure}

In the context of structure predictions and folding, CG interactive simulations exploit human creativity and constitute a technique complementary to massive computing.
While a regular MD of the system runs in the background, guided by our knowledge of the system, we can generate by manipulation highly different new conformations that a computer calculation may never reach in a finite time. 
These conformations can then be explored thoroughly and extensively by the enhanced simulations techniques presented earlier.
We have recently started to investigate this approach with HiRE-RNA, coupling the model and its simulation engine it with the MDDriver software \cite{Delalande2009} that allows to guide simulations interactively.
By connecting to a network socket, any device with a driver implementing the IMD protocol
can connect to the running HiRE-RNA simulation, and inject user forces to alter the simulations.
We used both the VMD \cite{Humphrey1996} and UnityMol \cite{Lv2013} programs to drive our simulations,
using either a mouse or a haptic device.
Interactive simulations allow us to easily fold and unfold hairpins \cite{Sterpone2014} (see figure~\ref{fig_interactif}) and to probe the stability of more complex structures.
The software we developed is currently being used also as a teaching tool in university courses. 
While the students benefit from the virtual reality experience of manipulating a molecular structure, we test the ability of interactive HiRE-RNA to solve folding problems by creativity.

\section{Benchmarking results} \label{sec_bench}
In this section we are going to discuss some of the results achieved by the various prediction methods, trying to draw comparisons where possible.
We'll start by discussing the 3D prediction competition set up by E. Westhof in 2012, to which bioinformatic methods mentioned in section \ref{sec_bioinfo} participated, together with the \textit{ab initio} method iFoldRNA.
The most recent \textit{ab initio} methods did not take part in the competition and are harder to compare on specific systems given the novelty of their codes.
We will illustrate results from their most recent publications, gathering similar systems.

\subsection{RNA-puzzles}
In 2012 the first RNA Puzzle competition was launched to benchmark prediction tools \cite{Cruz2012}, in the same spirit of what is done for proteins in the  CASP competition \cite{Moult2005}.
Different research groups attempt to predict RNA structures of molecules for which the experimental structure has been determined, but not yet published.

The sequences of three RNA were provided as challenge.
The first structure was a dimer (PDB ID 3MEI \cite{Dibrov2011a}) with symmetric sequences for the two strands, but for which
the crystallized structure displayed two asymmetrical internal loops.
The second structure was a square, composed of four duplexes, each of them
with the same inner and outer strands (PDB ID 3P59 \cite{Dibrov2011}).
The 3D coordinates of the inner strands were provided, leaving the outer strands
to be predicted.
The third structure was a riboswitch with a three way junction, PDB ID 3OWI \cite{Huang2010}.
The sequence of the crystallized structure had been modified at one loop,
compared with the sequence given to the contestants.
\begin{figure}[tb]
  \centering
  \includegraphics[width=0.4\textwidth]{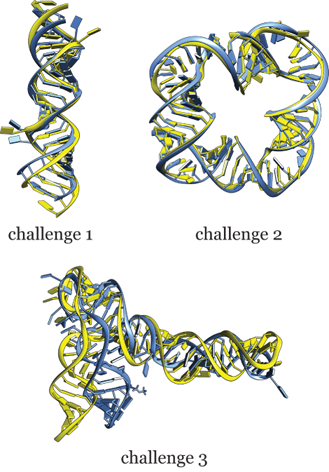}
  \caption{\label{fig_RNA_puzzle} Structures of the three challenges of RNA puzzle 2012.
The native structure in blue and the best prediction in yellow.}
\end{figure}

Other than the methods already discussed (Vfold, FARNA, MCFold, iFoldRNA), three other groups took part in the competition, for a total of seven participants.
The group by Bujnicki used the programs ModeRNA, based on sequence homology,  to obtain a tentative structure from known structures of similar sequence content,
and SimRNA to refine the structure through a 3-particles coarse-grained model and inverse Boltzmann potentials \cite{Rother2012}.
The group by Flores used the program RNABuilder performing Molecular Dynamics simulations in internal coordinates and rigidification of parts of the molecule \cite{Flores2010}.
Their force field consists of torques that act to fold the molecule according to restraints specified by the users and by stacking, which is the only interaction always present. 
Information for the restraints comes from experimental evidence, including sequence homology.
The group by Santa Lucia used the \textit{de novo} modeling module RNA123.
In this approach the secondary structure is predicted first and it is decomposed into constituent motifs such as internal loops, helices and hairpins, and the 3D structure is assembled putting together fragments from a motif library.

\begin{table}[tb] 
  \centering
  \begin{tabular}{ c | c | c | c | c | c }
    Challenge& Model & RMSD(\AA) & Rank & INF(\%) & Rank\\
    \hline & & & &\\
1 & FARNA		& 3.41		& 1 	& 0.93 		& 1		\\
1 & MC-Fold		& 4.06 		& 2 	& 0.89 		& 6 		\\
1 & VFold  		& 4.11 		& 3 	& 0.82 		& 4 		\\
1 & ModeRNA		& 4.66 		& 4 	& 0.81 		& 3		\\
1 & RNA123		& 5.67 		& 5 	& 0.84		& 5 		\\
1 & iFoldRNA	& 6.94 		& 6 	& 0.81		& 2 		\\
1 & RNABuilder	& -- 		& -- 	& -- 		& -- 	\\
\hline & & & &\\
2 & ModeRNA		& 2.30		& 1		& 0.81		& 4		\\
2 & FARNA		& 2.45 		& 2		& 0.86		& 1		\\
2 & iFoldRNA	& 2.54 		& 3		& 0.82		& 2		 \\
2 & VFold  		& 2.83 		& 4 	& 0.76		& 7		\\
2 & MC-Fold		& 2.98 		& 5 	& 0.78		& 6		\\
2 & RNABuilder	& 3.48 		& 6		& 0.79		& 5		\\
2 & RNA123		& 3.65 		& 7 	& 0.81		& 3 		\\
\hline & & & &\\
3 & VFold		& 7.24		& 1		& 0.74 		& 1		\\
3 & iFoldRNA	& 11.46 	& 2		& 0.71 		& 3		\\
3 & FARNA		& 11.97 	& 3 	& 0.73		& 2		 \\
3 & ModeRNA		& 12.19 	& 4 	& 0.62 		& 4		\\
3 & MC-Fold		& 13.70 	& 5 	& 0.59 		& 5		\\
3 & RNABuilder	& -- 	& -- 	& --		& --		\\
3 & RNA123		& -- 		& -- 	& -- 		& -- 	\\
  \end{tabular}
  \caption{Results of RNA-puzzles. For each challenge we report the RMSD of the best structure obtained by each method and the Interaction Fidelity Network (INF) of the structure with the best RMSD,
indicating the percentage of native base pairs recovered in the prediction.
Ranking of each method  is given separately for RMSD and INF and it refers to the performance of each method with respect to the others.
In the competition each group was allowed to submit more than one structure and results were ranked by best structure, independently of the method used.
Therefore the ranking we attribute here do not correspond to the ranking of structures given in the RNA-puzzle paper. Not all groups submitted structures for all challenges.}
\label{table_RNA_puzzle}
\end{table}
A complete discussion of the results can be found in the original RNA Puzzle paper.
In table \ref{table_RNA_puzzle} we give the summary of the best performances of the various methods.
The prediction for the duplexes are very close to the crystal structures for all methods.
Base pairs found in the experimental structure are recovered with high efficiency, which is not surprising since all base pairs in the dimer are
canonical (see also figure \ref{fig_RNA_puzzle}).
Of all three challenges, the best prediction were obtained for the square, where the coordinates of one strand were given as input data.
As expected helical regions were better predicted than loops.
The limits of all methods appeared clearly in the prediction of the riboswitch, for which all proposed structures deviate significantly from the experimental conformation.
From a detailed analysis of the different kind of interactions (canonical base-pairing, non-canonical, stacking), what appears to be most challenging is to predict the formation of non-canonical pairs. 
Only FARNA/FARFAR was able to recover a percentage (up to 60\% depending on the proposed structure) of non-canonical pairs, but the overall shape of the molecule was missed by the local reconstruction. 
The method that was able to better capture the overall shape of the molecule was VFold, which on the other hand missed completely the prediction
of non-canonical pairs.

Results of the RNA-puzzle competition highlight how for small, simple, RNA molecules we now have several well performing methods to predict the three-dimensional structure. 
However these methods only give access to a static picture of the structure and do not give insights on the dynamic of folding, on the thermodynamics, and do not account for the influence of external conditions such as ionic conditions or the possible presence of ligands. 
When challenged with RNA molecules of larger size and more complex architectures, such as the riboswitch, most methods give predictions far from the experimental structure and that for the most part do not recover the specific base-pairing network necessary for the molecule to hold its shape.

More recently, Xia tested his model with the same challenges posed by the RNA puzzle competition.
Coupling their model with secondary structure prediction algorithms providing 14 base-pairs, they were able to fold the riboswitch to the correct topology with an RMSD of 7\AA~ \cite{Xia2013}.
This is an important indication that the new generation \textit{ab initio} model, representing more realistically base-pairing and with the possibility of forming multiple pairs, begin to have prediction capabilities comparable to bioinformatic methods, with the added value of providing information also on the dynamics and thermodynamics of the molecule.

\subsection{Pseudoknots}
Predicting small pseudoknots is particularly challenging because of the tight configuration adopted by the molecule.
Small pseudoknots are stabilized extensively by stacking interactions, and base-pairing alone is not sufficient to hold them together.
Both OxRNA and HiRE-RNA, in its latest version, have been able to fold simple pseudoknots topology varying in length from 22 to 34 nucleotides.
It is interesting to notice that while the previous versions of our model were not able to predict such structures,
the more detailed description of hydrogen bonding and stacking introduced in HiRE-RNA v3 makes it possible fold tight pseudoknots as well.
OxRNA studied the thermodynamics of the experimentally well documented MMTV pseudoknot, composed of two WC stems of 6 base pairs and 5 base pairs respectively.
Their simulations recovered the double peaked melting temperature also found experimentally \cite{Sulc2014a}
that can possibly correspond to a higher temperature transition from unstructured strand to hairpin,
and a lower temperature transition from hairpin to pseudoknot.
Our investigation of the smaller 2G1W also support this behavior.
Starting from completely unfolded structures, REMD simulations predict the experimental fold as the most stable structure
at room temperature, and the corresponding specific heat curves also exhibit two separate peaks. 
Even though it was not possible at this stage to characterize the intermediate state between unfolded and pseudoknot, ST simulations showed the presence of non-negligible populations of distorted hairpins.
 \begin{figure}[tb]
   \centering
   \includegraphics[width=0.45\textwidth]{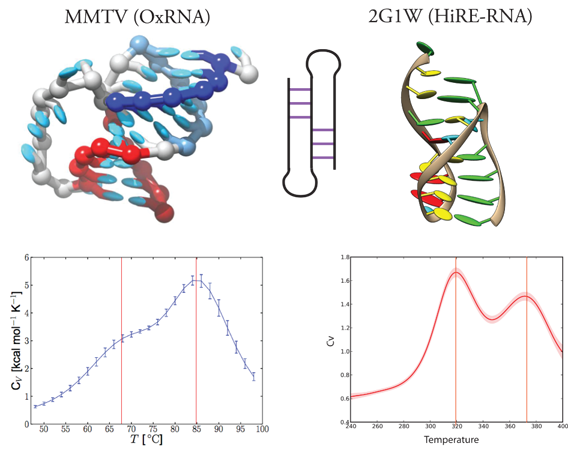}
    \caption{\label{fig_pkPMF} Top: Folded structures obtained by OxRNA and HiRE-RNA for the MMTV and 2G1W pseudoknot, respectively.
    Bottom: Specific heat as a function of temperature for both simulations,  both showing the presence of 2 peaks, corresponding presumably
    to the formation of a hairpin structure (high peak), followed by the formation of the folded pseudoknot (low peak). }
 \end{figure}

To highlight the importance of non-canonical pairings, we repeated REMD simulations on 2G1W
allowing only interactions between Watson-Crick sides.
The molecule is still able to reach the native state, but a large number of partially folded or misfolded states are missing,
pointing to the importance of considering non-canonical pairs even for molecules in which the native state consists purely of canonical base pairs.
Preliminary investigation seem to suggest that the possibility of forming transient non-canonical pairs opens new folding pathways, more easily interconnecting largely different conformations, involving rearrangements of the secondary structure.
This could be an important general feature of RNA, since many RNA molecules are known to be able to adopt structures with different architectures in response to the cellular environment and signals.

\subsubsection*{Triple helix folding}
HiRE-RNA was able to fold pseudoknots beyond the simple topology.
Indeed it was able to fold the triple helix of the telomerase RNP complex (2K96) \cite{Kim2008},  a compact structure exhibiting several multiple A triplets,  from extended configuration, and without any additional input other than the sequence.

\begin{figure}[bt]
 \centering
 \includegraphics[width=0.45\textwidth]{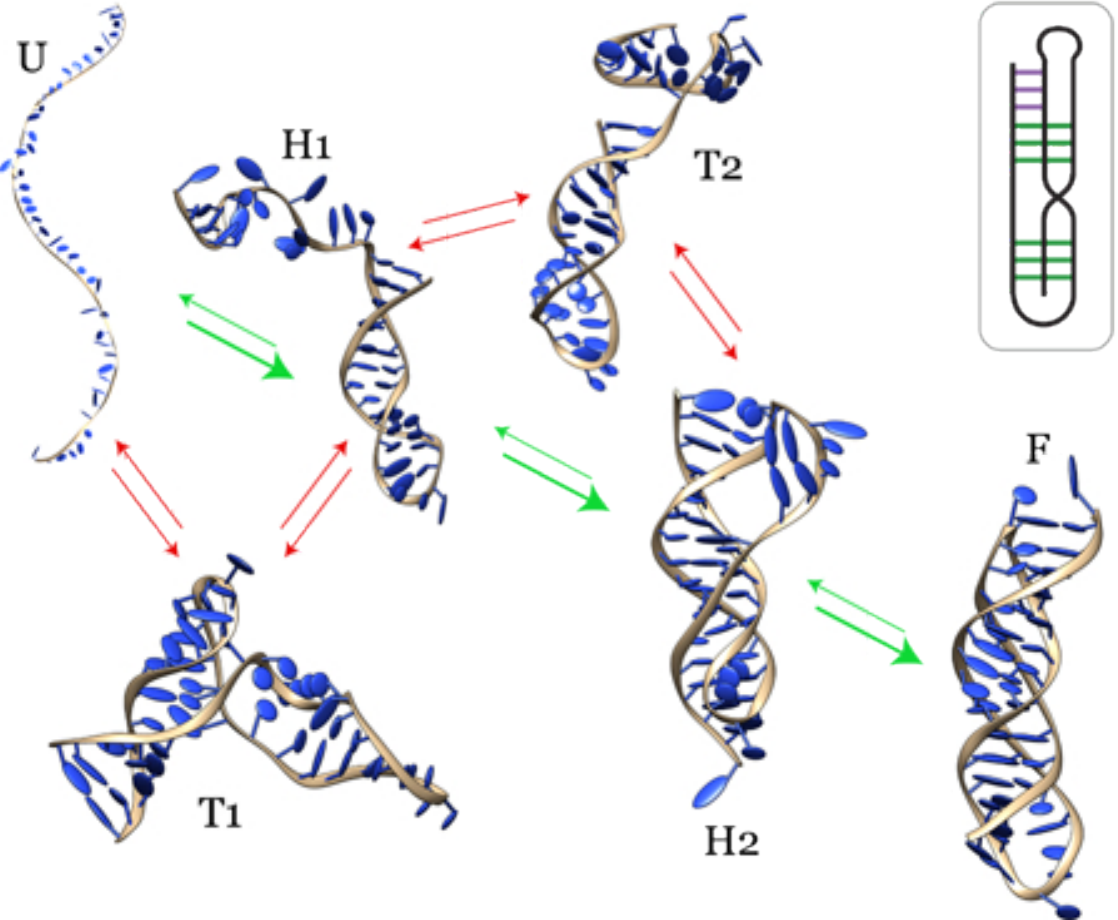}
 \caption{\label{fig_2K96} Conformational states and transition for 2K96 during a folding simulation.
 The molecule starts in a completely unfolded state (U) and rapidly transitions to a state where the WC helix is formed (H1).
At this stage the molecule can get trapped in misfolded states with alternative base pairings such as T1.
On a longer time scale the molecule transitions to a state where the pseudoknot is formed with two WC helices (H2)
and can get trapped into configurations where the dangling end folds back on itself such as T2.
Finally the remaining unpaired strand enters the groove of the first helix making triple contacts and giving rise to the full triple helix.}
\end{figure}

2K96 is a 49 nucleotides long structure composed of a WC hairpin with a dangling end that folds back on itself inserting into the groove  of the hairpin forming triple pairs, and locking into position by base-pairings with the hairpin loop .
In REMD, folding of the triple helix occurred in two steps.
In a short first phase the hairpin is formed, while the second phase is characterized by the insertion of the dangling end into the groove.
The correct topology was formed after 600ns REMD time with the molecule reaching a native RMSD of 7-8 \AA~.
After 1.2\(\mu\)s, the full NMR base pair network was reached and the RMSD lowered to 4.1\AA~ with respect to the NMR structure (figure \ref{fig_2K96} ). 

To our knowledge HiRE-RNA is the only model that was able to fold a structure heavily stabilized by multiple base pairs,
a result that is even more significant considering no base-pairing biases or other constraints has been used.
This is an encouraging result for all coarse-grained \textit{ab initio} models, as it shows that when the correct ingredients are put into the effective theory,
on one hand predictive capabilities are high, and on the other the model provides some key elements for our understanding of the folding mechanism.

\section{Coupling predictions and experimental data} \label{sec_exp}
A good \textit{ab initio} model should in principle obtain 3D predictions starting from the sequence knowledge only.
This is the case for HiRE-RNA and for Xia's model for relatively small RNA molecules.
However, the cost of exploring all possible pairings becomes quickly unmanageable as the system size increases.
On the other hand, experimental information is often available on the systems of biological interest, and even if minimal, it can be of great use to help models converge to a
sound prediction.
As we have seen for bioinformatic models, secondary structure predictions can also be integrated into the prediction pipeline with good results for molecules with simple architectures for which 2D models are reliable.

Experimental information fall into two categories: local, high-resolution, data, which include all informations on the spatial proximity of parts of the molecules, typically base-pairing,
that can be obtained by NOESY resonances from NMR, exposed surface data with SHAPE \cite{Weeks2010}, or pairs obtained by secondary structure predictions \cite{Mathews2007};
and low-resolution, global, data  giving information on the shape of the molecule, that is obtained by Small Angle X-ray scattering (SAXS) \cite{Putnam2007,Parisien2012} and Cryo-EM \cite{Mitra2006,Whitford2011,Lindert2013}.
Different information can be integrated by prediction models in various ways.
In bioinformatic models, Vfold for example, local base-pairing data can be used to give a complete secondary structure of the molecule which is then turned into a 3D scaffold constituting the starting point of a refined prediction. 
In \textit{ab initio} models, where the folding process is simulated, base-paring can be introduced as a set of constraints that are implemented at the same time as the molecule folds.
This is the strategy we have adopted with HiRE-RNA to fold a riboswitch, for which we are able to recover the overall topology and secondary structure using only 3 base-pairs constraints.
Global, low-resolution, data, is less direct to include into models, but it can easily be used as filter for proposed structures.

\subsection{Local constraints}
In \textit{ab initio} simulations, where a force field is defined, local constraints can be easily implemented in the form of additional potentials, exerting a force on some particles of the system.
Adding this information in a simulation can lead to dramatic improvements in its ability
to reach the native state by eliminating or hindering the exploration of a large number of possible conformations
now incompatible with the imposed data.

Depending on the functional form of this additional potential, different strategies can be adopted.
A soft pairing potential, dying off at long distances, provides additional stability to base pairing once the pair is formed or nearly formed,
but it is not be able to drive the molecule to the specific conformation starting from a completely different state. 
A hard potential, extending to long range, such as an harmonic potential or a potential having a linear behavior at large distances, is able to drive the molecule
to form the requested pairs from unfolded configurations. 
However, care must be taken in the way this potential is added in the simulation.
While the observed base pairs may well be present in the folded structure, if applied too quickly or too strongly,
the constraints can dominate the molecule's behavior resulting often in entangled states,
where the natural behavior of the unconstrained molecule is violated:
forcing the constraints at all times in the simulation may stabilizes partially folded states,
thereby significantly slowing down the folding process.

With HiRE-RNA we decided to adopt a long-range potential, harmonic over short range and linear beyond a 4\AA~ cutoff, to impose a set of preassigned base-pairs.
The switch to a linear potential at larger distance allows to modulate the maximum force present in the system, so that even for large deviations from the target constraints values,
the system's forces can be numerically integrated without problems.
To prevent the system from being locked by the constraints in misfolded intermediates,
the potential is modulated in time so that the molecule goes through phases when the constraints are active and phases
when the molecule can relax unconstrained to re-establish its natural topology.
With HiRE-RNA the use of just a few local constraints dramatically improved the success in folding.

\subsubsection*{Riboswitch}
\begin{figure}[tb]
 \centering
 \includegraphics[width=0.35\textwidth]{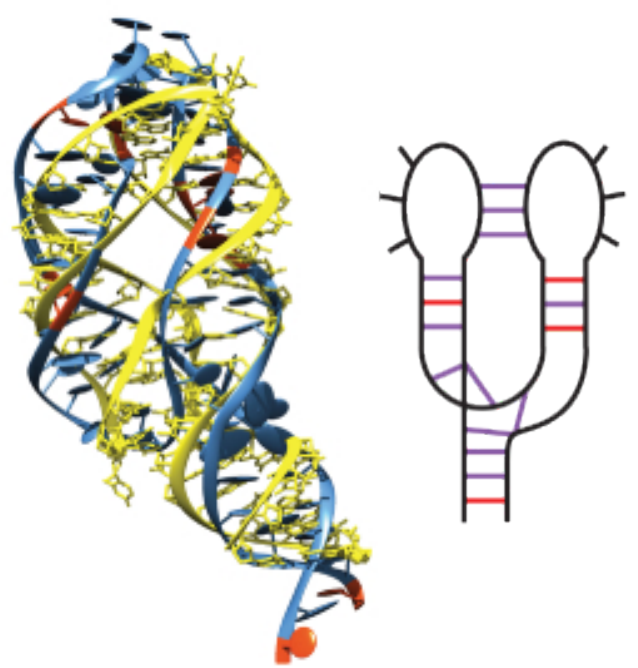}
 \caption{\label{fig_1Y26}  Predicted structure of 1Y26 superposed to the experimental configuration (yellow) next to a schematic representation of the secondary structure and tertiary contacts (right). The three imposed constraints are shown in red in both 2D and 3D representations.}
\end{figure}

As we saw already in the discussion of RNA puzzle results, folding large structures is challenging because of the complex architectures they can adopt, with many possible alternative secondary structures.
We have used constrained REMD simulations to obtain the structures of the 79 nts adenine riboswitch (1Y26) \cite{Serganov2004}.
In its NMR state with an adenine ligand the riboswitch adopts a Y shape with the two upper stems binding through kissing loops.
To test whether we could fold this large molecule, we imposed three secondary structure constraints taken from the experimental structure, one on each of the helices.
Imposing the time-dependent potential, both simple MD (at 300K) and REMD were able to recover the overall organization of the kissing loops with RMSD of 7-8\AA~ (figure \ref{fig_1Y26}).
The structure is not yet determined with high accuracy, but our results are comparable with the best results obtained by other techniques for ribozymes of similar size.

How to determine the optimal number of constraints to make the simulation converge quickly to the experimental configuration remains an open question.
Our example using HiRE-RNA shows that a very limited number of constraints can suffice to obtain the overall shape of the molecule, but clearly a larger number of constraints could improve precision if occurring in different parts of the molecule.
On the other hand, constraints occurring on the same helix do not provide much additional information in the context of folding since when a base-pair is imposed in a simulation
near by pairs form spontaneously. 
Many constraints could however trap the molecule in entangled states if not applied carefully. 
Future work will address the question of what is a good set of local constraints, both in terms of how many are needed, and in terms of their distribution on the global architecture.

\subsection{Global constraints}
SAXS experiments are now becoming common to investigate biomolecular structures since they have the advantage of leaving the molecule unperturbed in its own environment.
An X-ray beam is shined for several minutes through a solution containing the molecule, water and ions, at room temperature, and forward scattering is recorded.
The collected data is an intensity curve as a function of exchanged momentum $q$.
Because the molecule is free to move in solution and it is observed over an extended time, the intensity curve is the result of a spatial and temporal average, which could be deconvoluted to give complete structural information only in the limit of $q$ going to infinity.
In practice only the forward scattering has enough intensity to give a clear signal, limiting useful data only to low $q$ values (forward scattering).

Once the atomic structure of a molecule is known, from high-resolution experiments or as a model from simulations, SAXS intensity curves can be calculated explicitly through Debye formula \cite{Debye1915}, involving a double summation of atomistic scattering factors of all atoms of the molecule and of the solvent surrounding it. 
For large molecules, as it is the case for proteins, RNA, and for protein-RNA complexes, these calculations become rather demanding, especially if one needs to repeat them often, as it would be the case for a comparison of a simulation trajectory with experimental data where one would need to compute SAXS curves at regular intervals for different simulated configurations.
A way of circumventing this problem comes from the consideration that SAXS data are intrinsically at low-resolution and it is therefore natural to couple them with coarse-grained models. 
Debye's theory can be generalized to grains, instead of atoms, by averaging atomistic form factors \cite{Harker1953,Yang2010}. 
This technique has been applied successfully for proteins and it is starting to be investigated for RNA molecules as well \cite{Yang2009,Putnam2013}.
A coarse-grained molecular representation was used to perform SAXS comparisons of structures predicted through MC-Fold and MC-Sym \cite{Yang2010,Parisien2012}.
Theoretical SAXS curves are computed using a 2-bead model (backbone-base) and compared to experimental intensities curves to filter structures that satisfy the experimental data.
An important aspect of the process is the treatment of the solvent since SAXS intensity curves depend on the contrast of the molecule's scattering with the solvent's scattering.
In the MC-Fold pipeline, the molecule is hydrated by a layer of dummy water molecules and ions.
Application of this method to a tRNA, to the P4-P6 intron domain, and to an RNA dimer, show that filtering with SAXS data is able to select native-like folds.
With a similar  approach, Xia and coworkers also set up a pipeline where models proposed by simulations with their coarse-grained 5 particles model are filtered by comparison with SAXS amplitudes after having been converted into atomistic structures. 
\begin{figure}[t]
 \centering
 \includegraphics[width=0.45\textwidth]{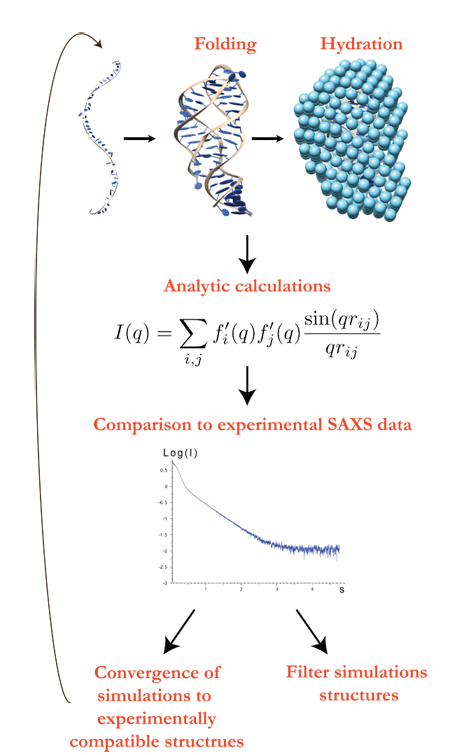}
 \caption{\label{fig_SAXS} Pipeline of coupling SAXS intensity curves and prediction tools or simulations.
Step 1: the prediction method is used unbiased to generate plausible structures,
Step 2: structures are hydrated to include the critical water layer around the molecule affecting scattering curves,
Step 3: analytic calculations of SAXS curves are computed using Debye's formula and compared with experimental data (Step 4).
Step 5: structures can be filtered according to their fit with experimental data (Mc-Fold) or simulations can be biased (HiRE-RNA) looping back to Step 1.}
\end{figure}

With HiRE-RNA we want to go one step further than simply filtering simulation results, and we are testing the possibility of integrating SAXS experimental data "on the fly" to make the simulation converge toward structures compatibles with the low-resolution data (work in progress). 
Thanks to the important reduction in the number of particles, SAXS calculations can be performed often during the course of a simulation without significantly slowing it down.
The fit between SAXS theoretical curves and experimental data is integrated as an additional term in the potential computed only with a preset frequency, with times large enough so that the molecule has adopted a significantly different configuration, that can indeed give rise to a different scattering intensity curve, but short enough so to lead the simulation to converge to the experimental data.
As it is the case for MC-Fold, hydration is a critical step.
For this purpose we introduce a hydration layer of coarse-grained water molecules all around the RNA.
The main steps of integrating SAXS data into simulations is highlighted in figure \ref{fig_SAXS}.

\section{Open challenges} \label{sec_chall}
The physical description of RNA folding process and the prediction of equilibrium structures is a field still in its early stage.
As it was shown in the previous discussion, properly accounting for base-pairing and stacking is challenging when one leaves the atomistic description, but it is 
a necessary step to study RNA molecules of sensible sizes.
Other interactions are undoubtedly relevant for RNA folding and have yet to be integrated into \textit{ab initio} models.
Two main additional challenges come from the interaction of the molecule with the surrounding ions and the interaction of bases with the phosphate groups.

\subsection{Ions}
Accounting for ions is a challenge for all simulations, atomistic or coarse-grained.
In atomistic simulations, with fixed partial charges, the effect of ions is poorly represented because of the lack of polarization \cite{Allner2012}.
In coarse-grained models, ions are typically not represented explicitly and their effect is taken into account only indirectly.
It is clear that neither approach is satisfactory at this stage \cite{Bowman2012}.

In RNA one can distinguish three different roles of ions.
The first long-range role is that of counter-ions, present to reduce the electrostatic repulsion of phosphate groups.
The second role is ion condensation around the molecule, constituting a dynamical layer of additional charge on the molecule's surface.
The third role is that of structural ions, where ions are observed in specific positions in crystal structures.
Most \textit{ab initio} models account for an electrostatic interaction between phosphate groups that can take implicitly into account the role of counter-ions, usually modeled though a Debye-Huckel potential, but lack a description of the other two aspects.

Xia and co-workers have shown the impact of introducing structural ions on a small pseudoknot \cite{Xia2013}.
For 1L2X the effect of magnesium ions was accounted for by imposing distance restraints between phosphates known experimentally to be in contact with the ion, imposing and stabilizing the shape of the ion pocket throughout the simulation. 
Their results showed how indeed the structure can be sensibly ameliorated when structural ions are kept into account.

With the example of HiRE-RNA we want to investigate the effect ion condensation, and possibly structural ions, and show how a simulation in the presence of a few explicit ions differs qualitatively from a simulation where an exclusively implicit ion description is used, underlying the importance of a correct ion representation to understand the structure of RNA and the dynamics of ion interactions \cite{Draper2008}.
These very preliminary results give us some insights on how ion condensation and structural ions could be considered in \textit{ab initio} models.

\subsubsection*{Short range RNA-ions interactions}
Two challenges are associated with short-range RNA-ions interactions. 
First, the charge density in those area is high enough
that most implicit solvent methods, such as Generalized-Born and Poisson-Boltzmann solvers, exhibit divergent results \cite{Kirmizialtin2012a}.
Also, those methods are ill-suited to represent accurately the strongly constrained geometry that are caused by those interactions.
Mg\(^+\) ions, for instance, are known to have an octahedral first coordination shell,
constraining the surrounding RNA backbone when tightly bound in an RNA structure.
Second, the precise geometries seen around those ions also suggests that traditional terms used to represent the nonbonded interactions
between particles in force field, such as the Lennard-Jones or Coulomb term, may not be adequate either
despite progress in their parametrization\cite{Allner2012},
since they have uniform value for a given radius distance, and would not encompass the important angular aspect of the interaction.
\begin{figure}[tb]
 \centering
 \includegraphics[width=0.4\textwidth]{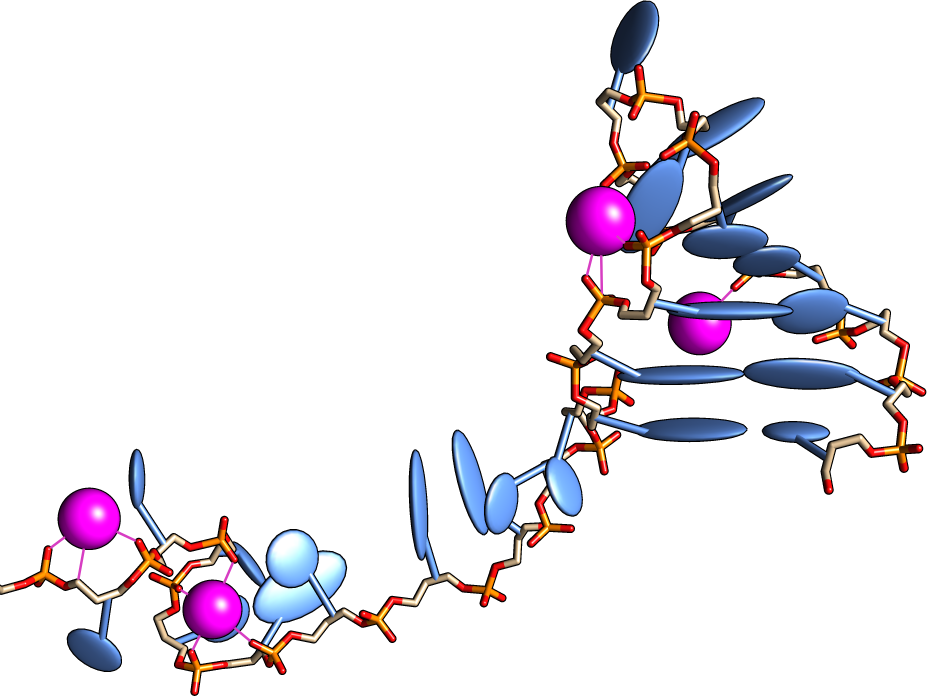}
  \caption{\label{HiRE_ion} HiRE-RNA structures with strong bending of the backbone caused by the presence of those ions.}
\end{figure}

To study the effect of magnesium ions in our simulations, we started supplementing the HiRE-RNA model
with an explicit description of magnesium ion particles.

\begin{figure}[tb]
 \centering
 \includegraphics[width=0.4\textwidth]{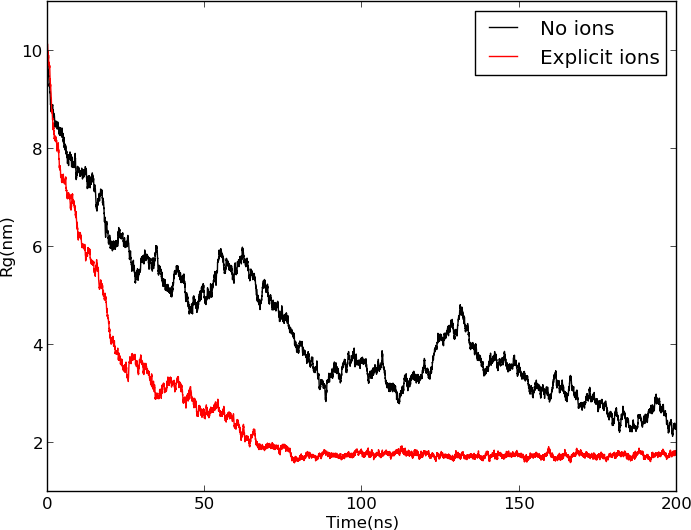}
 \caption{\label{fig_ion} Evolution of the radius of gyration for a RNA during simulations started from an extended conformation. The presence of magnesium ions greatly increase the speed of this initial compaction.}
\end{figure}

We started with a classical point-charge description of the charged ion,
in order to observe which phenomenon would be captured by such a model.
The most striking result of these simulations, illustrated in figure~\ref{HiRE_ion},
is the strong bending of the backbone induced by the ions.
Even at very low concentration, the addition of ions increases the rate of formation of the local hairpins by at least an order of magnitude, favoring compaction.
With a concentration of approximately 0.1 mM, a large increase in the speed of compaction can be seen compared with a simulation without ions (figure~\ref{fig_ion}).
This is in agreement with experimental data, where the presence of magnesium ions greatly improves folding rates, and can shift the equilibrium of RNA molecules toward folded structures \cite{Onoa2004}.

\subsection{Base-Phosphate interactions}
\begin{figure}[tb]
 \centering
 \includegraphics[width=0.4\textwidth]{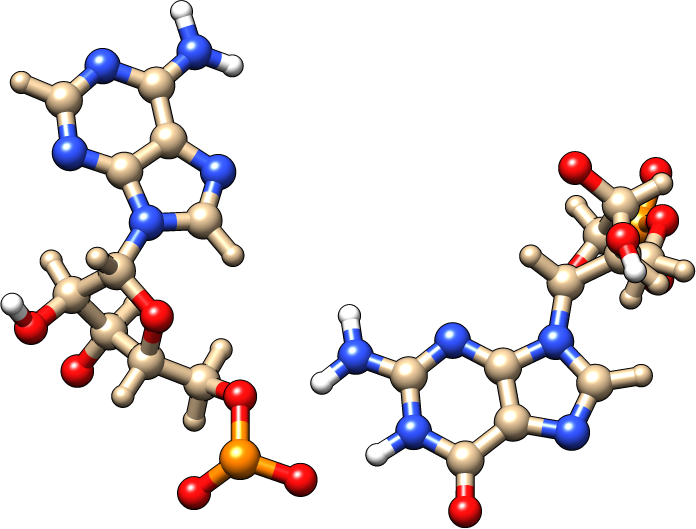}
 \caption{\label{fig_BPh} Base-phosphate interactions of bases G and A of the sarcin/ricin loop, with two hydrogen bonds, classified as 4BPh in \cite{Zirbel2009}.}
\end{figure}

It has recently been recognized that hydrogen bonding between bases and the oxygens of the phosphate groups is a common interaction in RNA molecules of complex architectures \cite{Zirbel2009}, for example in the S-turn motif present in the sarcin/ricin loop,
a part of the ribosomal RNA that forms an essential binding site for elongation factors \cite{Correll1999} (figure \ref{fig_BPh}).
Currently no structure prediction method, bioinformatic nor physics-based, explicitly accounts for these interactions which are also miss-represented by atomistic non-polarizable force fields.
A statistical investigation of the NDB, followed by quantum calculations, has shown that each base can form stable interactions with the phosphate group. 
A classification has been proposed for the most common base-phosphate motifs based on the base's hydrogen fixed by the phosphate's oxygen.
For each base several possible fixation point are available and quantum calculations were able to investigate the relative stabilities of these bonds in implicit solvent.

In a model such as HiRE-RNA where the phosphate group is corresponds to a bead, and where bases are described with sufficient accuracy to distinguish their sides, and therefore differentiate between possible fixation points, a base-phosphate potential term could be easily introduced.
As it is the case for all other types of interaction, one would need to find the appropriate form for this potential as well as the relative weight of its contribution with respect of all other interactions.
The statistical analysis of existing structures could be translated into a statistical potential and quantum mechanics calculations be used as weights.
However, consistently with the development of the other potentials of HiRE-RNA, we are considering designing a phenomenological potential derived directly from quantum calculations, and then adjusting its weight through the comparison with existing base-phosphates motifs.

\section{Perspectives} \label{sec_future}
RNA structure prediction is a young field where much is yet do be done. 
Bioinformatic tools are in continuous evolution to better keep into account experimental data in their predictions, as the example of MC-Fold and SAXS data clearly shows, but it is  physics-based \textit{ab initio} models that have the largest margins of improvement and where much has yet to be accomplished.
High-resolution models such as Xia's and HiRE-RNA have shown their potential as proof of principle and have now to be put to the test in a variety of real systems.

As a start, existing models need to be used for predictions of systems beyond benchmark molecules. 
It is only with "real life" tests that we will learn their strengths and their weaknesses.
This is an essential step in the development of any model. 
The fact that physical models make predictions also on the dynamics and thermodynamics of the molecule gives access to the comparison with experimental data other than crystal or NMR structures.
Non-structural data contribute to our understanding of the relevant terms in folding helps ameliorating the force-fields.

A second line of development is directed toward ameliorating the description of the system to include critical elements left out at first.
A successful model should keep into account the critical role of ions as well as base-phosphate hydrogen bonds, implying that new force-fields will have to be designed and parametrized.

A third and more challenging development is to account for the environment and for the conditions under which folding occurs.
Unless we limit ourselves to folding experiments in vitro where the molecule has been completely isolated from its natural context, predicting a functional RNA structures implies understanding, and possibly controlling, all other processes surrounding it.
RNA molecules can interact with ligands, that for example for riboswitches can drive the formation of one or the other of the possible structures \cite{Reining2013}, form complexes and interplay with proteins, fold concurrently with their transcription from DNA (co-transcriptional folding) \cite{Zhang2011}, and are subject to the complex cellular machinery of degradation which filters out misfolded structures.
It would be presumptuous, at this stage, to think that we can tackle all of these questions with the tools currently at our disposal, but there are some points worth considering before becoming overspecialized in making predictions for isolated molecules.

\subsubsection*{Co-transcriptional folding}
\begin{figure}[tb]
 \centering
 \includegraphics[width=0.30\textwidth]{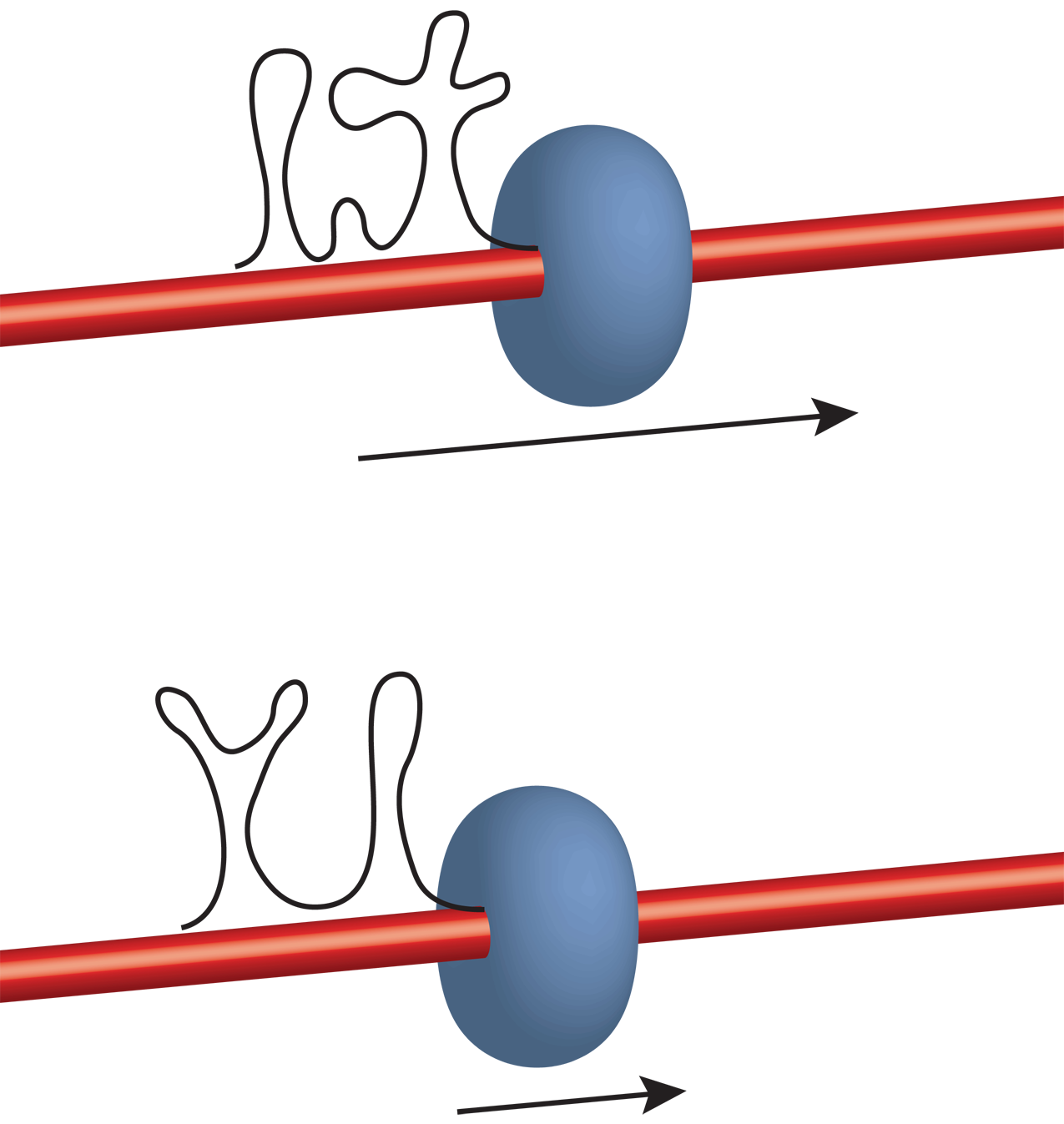}
 \caption{\label{fig_fdyn} Cartoon of co-transcriptional folding. As the RNA molecule (black string) is transcribed from the DNA (red tube) thanks to the action of the polymerase (blue donut), it starts folding. The folded structure is in this case the results of a dynamical process that depends also on the speed (arrow) and pauses of the polymerase along the DNA.}
\end{figure}

One aspect that we can easily access with simulations is co-transcriptional folding vs. free folding.
In MD simulations where we have access to the dynamics of folding, we can release the degrees of freedom of the molecule a bit at the time and simultaneously let the unfrozen portion of the molecule fold. 
Such a procedure can be implemented immediately for any model implying MD simulations and it is currently under investigation with HiRE-RNA.
It is then straightforward to mimic some basic biological processes as for example the speed at which the polymerase produces the RNA  \cite{Schroeder2002} and that would change the speed of release of the frozen degrees of freedom in a simulation (figure \ref{fig_fdyn}).
Without adding too much difficulties, one could also go one step further and explicitly include an DNA/RNA double helix with the dangling RNA released over time and simultaneously folding.
Such studies will help us understand the context in which folding take place and the relative importance of the different phenomena.

\subsubsection*{Post-processing}
\begin{figure}[b!]
 \centering
 \includegraphics[width=0.4\textwidth]{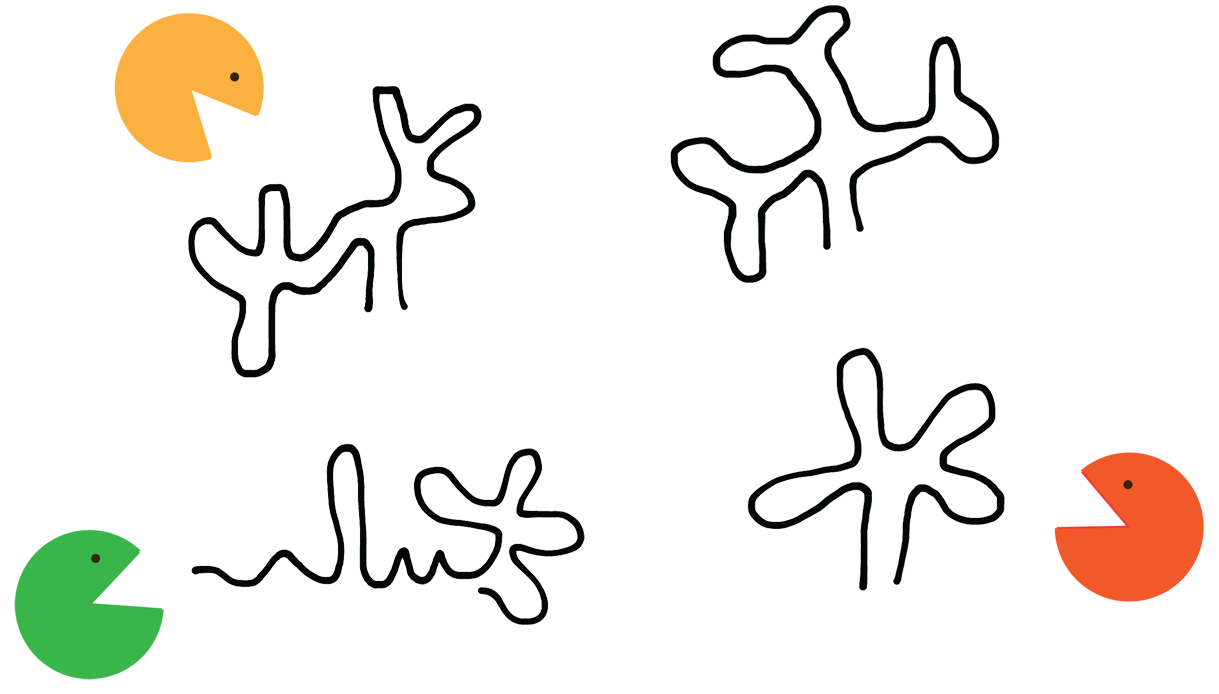}
 \caption{\label{fig_pacman} Cartoon of the filtering mechanism of the ribonucleases (pacman) that are able to recognize and degrade misfolded structures, leaving only functional molecules to be active in the cell.}
\end{figure}

As it is shown by riboswitches, RNA molecules can adopt alternative structures of similar energies separated by hight barriers, involving the reorganization of the secondary structure and therefore the formation and breaking of many base-pairs. 
Such barriers are unlikely to be overcome simply by thermal energies at room temperature.
In biological systems, it is more probable that during folding the molecule gets trapped in stable states far from the native structure from which it can't escape on its own, and what we observe as functional structure is the product of a selection process carried out by RNA degradation factors (figure \ref{fig_pacman}).
The degradation machinery is itself a complex system, out of our modeling reach, but coarse-grained models coupled to enhanced sampling techniques can generate a large variety of alternative structures and can shed light on the plurality of possible configurations that can co-exist at different stages of the cell's life and that can be a pool of new functional structures if environmental conditions change and the need presents.
Models such as HiRE-RNA, Xia's or OX-RNA, can all give access to large sets of very different structures and investigate their structural properties and relative equilibrium.

\subsubsection*{Protein partners}
RNA molecules often work in conjunction with proteins in an intricate functional network.
If the structure of the RNA and of the protein are known, protein-RNA docking programs can give insights on the possible interactions between the two partners under the assumption that neither the protein nor the RNA substantially modify their conformations from unbound to bound \cite{Lang2009}. 
This is the case for some complexes, but not for all. 
It is well known that formation of many complexes involves a substantial rearrangement of the protein or of the RNA or both.
Examples of structural reorganizations upon binding in complexes are ubiquitous and go under the name of "induced fit" \cite{Williamson2000,Schwalbe2007}.
Because of the large size of a complex, once again atomistic simulations are not suited for the study of systems where large scale structural changes are involved, but coarse-grained models could stand up to the task.
Currently a fully-flexible coarse-grained model including a description for both proteins and nucleic acids  and for their interactions does not exist.
The widely known Martini force-field \cite{Marrink2013} can describe simultaneously proteins, lipids and nucleic acids, but it has to impose secondary structures.
It can therefore accommodate for molecular flexibility, but not for induced large-scale rearrangements.

HiRE-RNA has a protein coarse-grained model parter named OPEP \cite{Sterpone2014} that has been successfully used over the last 10 years to address protein folding, in particular in the context of amyloid fibers formation in relation to Alzheimer's disease. 
At its present stage OPEP can take into account many features specific of amino acid interactions such as hydrogen bond formation to give rise to secondary structures, salt bridges, and hydrodynamic effects. As simplified model with only 6 particle per amino acid, it can study large scale rearrangements of single molecules as well as the interplay of several thousands molecules at once.
HiRE-RNA and OPEP have been developed in parallel on the same simulation engine with the goal of one day bringing the two together into one model able to describe the behavior of co-existing proteins and RNA, allowing full flexibility for both.
Now that HiRE-RNA is converging into a stable version and has given proof of being able to address RNA folding correctly, we are starting to work on developing an interface force field that will take into account occupied volumes, electrostatic interactions, but also specific interactions between bases and amino-acids, making use of the experience we have developed to treat the details of base interactions.

\begin{acknowledgments}
The authors wish to thank the French ANR LABEX DYNAMO ANR-11-LABX-0011 for equipment support, P. Stadlbauer and J. Sponer for kindly providing the G-quadruplex image, P. Sulc, T. Ouldridge and J. Doye for providing the images of the MMTV pseudoknot, M. Baaden and S. Doutreligne for the interactive simulation image, B. Caron and Y. Laurin for their contribution in the development of HiRE-RNA during their Master internships.
\end{acknowledgments}

\bibliographystyle{ieeetr}
\bibliography{publis}

\end{document}